\documentclass[11pt,a4paper]{article}
\pdfoutput=1
\usepackage{jheppub}
\usepackage{xcolor}
\usepackage{latexsym,amsfonts,amsmath,amssymb}
\usepackage{bbm}
\usepackage{empheq}
\usepackage{graphicx}
\usepackage{color}
\usepackage[compatibility=false]{caption}
\usepackage{subcaption}
\usepackage[normalem]{ulem}
\usepackage{comment}
\usepackage{url}
\usepackage{slashed}
\usepackage{tabu}
\usepackage{multirow}
\usepackage{extarrows}
\usepackage{amsmath}
\usepackage{mathtools}
\usepackage{cancel}
\usepackage{amsxtra,graphics,epsfig,bm,tikz,xfrac,lscape}
\usetikzlibrary{decorations.pathmorphing}
\usetikzlibrary{decorations.markings}
\usetikzlibrary{arrows, decorations.markings, calc, fadings, decorations.pathreplacing, patterns, decorations.pathmorphing, positioning}

\usetikzlibrary{positioning,shapes}
\usetikzlibrary{chains}
\usetikzlibrary{arrows,fit,decorations.pathreplacing}
\tikzstyle{every picture}+=[remember picture]
\tikzstyle{na} = [baseline=-.5ex]

\title{
Moduli-dependent one-loop entropy of hyperbolic BPS black hole in AdS$_4$
}

\author[a]{ Imtak Jeon}
\affiliation[a]{School of Science, Huzhou Normal University, Huzhou 313000, Zhejiang, China}

\author[b]{, Alokananda Kar}
\author[b]{, Binata Panda}
\author[c]{, and Anowar Shaikh}
\affiliation[b]{Department of Physics, Indian Institute of Technology (Indian School of Mines) Dhanbad, \\Jharkhand 826004, India.}
\affiliation[c]{Department of Physics, Indian Institute of Technology Bombay, Mumbai 400076, India.}

\emailAdd{imtakjeon@gmail.com}
\emailAdd{alokanandakar@gmail.com
}
\emailAdd{binata@iitism.ac.in
}
\emailAdd{anowarfizik@gmail.com
}

\abstract{
We study one-loop logarithmic corrections to the entropy of static hyperbolic BPS black holes in asymptotically AdS$_4$ spacetime. Our analysis is carried out in a consistent real-scalar truncation of ${\cal N}=2$ Fayet-Iliopoulos gauged supergravity specified by the prepotential $F=-i X^0 X^1$, which corresponds to an Einstein-Dilaton-Maxwell theory with a nontrivial scalar potential. In this model, the classical BPS attractor mechanism exhibits flat directions, leaving scalar moduli on the black hole horizon unfixed, while the Bekenstein-Hawking entropy depends only on the charges. We show that the resulting one-loop correction to the black hole entropy acquires a nontrivial dependence on the horizon moduli and induces an effective quantum potential that dynamically stabilizes them at a preferred value. Our results provide an explicit and concrete realization of quantum lifting of classical attractor flat directions in gauged supergravity.
}

\begin{document}

\maketitle

\section{Introduction}
Logarithmic corrections to black hole entropy provide a powerful infrared window into the microscopic structure of quantum gravity. Although they arise from one-loop fluctuations of low-energy effective fields, their coefficients are protected under renormalization and therefore encode robust information about the underlying quantum microstates. Thus, logarithmic corrections offer a rare observable that is simultaneously computable within effective field theory and sensitive to ultraviolet data.

This idea was first formalized by Sen through the introduction of the quantum entropy function (QEF) \cite{Sen:2008vm,Sen_2009}, which defines the entropy of extremal black holes in terms of a Euclidean path integral formulation \cite{Gibbons:1976ue,Hawking:1978jz} with boundary conditions dictated by the black hole near-horizon geometry,
\begin{equation}\label{eq:QEF}
    Z_{\text{AdS}_2} = \int_{\text{AdS}_2} D\Psi \, e^{{\cal S}[\Psi]
    }
\,, \qquad {\cal S}[\Psi] = -S_E - i q_I \oint A^I\,,
    \end{equation}
    where $S_E$ is the Euclidean action. 
In this formulation, the exponent ${\cal S}[\Psi]$ defines the entropy function~\cite{Sen:2008yk}, whose extremum reproduces the classical Bekenstein–Hawking entropy \cite{Bekenstein:1973ur,Hawking:1975vcx}, while one-loop fluctuations around the classical saddle generate logarithmic corrections of the form
    \begin{equation}\label{Bekenstein+Log}
     {\cal S}_{\text{QEF}}=   
     {\cal S}^{\text{class.}}_{\text{BH}}+ C \ln \frac{A_H}{G_N}  +\cdots \,.
    \end{equation}

These logarithmic corrections have been successfully computed for a wide class of supersymmetric black holes in asymptotically flat spacetime and shown to agree with microscopic state counting in string theory \cite{Banerjee:2010qc,Banerjee:2011jp,Sen:2012kpz, Sen:2012cj} (see \cite{ Sen:2014aja} for a review). Over the past decade, this program has been extended to a variety of gravitational systems, including one-loop tests of AdS$_4 \times S^7$ supergravity \cite{bhattacharyya2013onelooptestquantumsupergravity,Bobev:2023dwx}, 
black holes in asymptotically AdS spacetime \cite{Jeon_2017, Liu:2017vll, Liu:2017vbl, PandoZayas:2020iqr, Gang:2019uay, Benini:2019dyp, David:2021eoq,Karan:2022dfy,Karan:2024gwf}, and (non-)extremal black holes \cite{Sen:2012dw,Banerjee:2021pdy, Charles:2015eha,Bhattacharyya:2012wz,Castro:2018hsc,Karan:2019gyn,Karan:2020njm,Karan:2021teq}. Further nontrivial consistency checks have been achieved through gravitational indices \cite{Iliesiu:2021are,H:2023qko,Anupam:2023yns}, as well as near-extremal analyses using small-temperature expansions \cite{Iliesiu:2022onk,Banerjee:2023quv,Banerjee:2023gll,Modak:2025gvp}. 

A particularly interesting arena for logarithmic corrections is provided by black holes in asymptotically AdS spacetime. On the microscopic side, the statistical interpretation of supersymmetric AdS black holes—first constructed within gauged supergravity \cite{Caldarelli:1998hg,Cacciatori:2008ek,Gutowski:2004ez,Cacciatori:2009iz,Hristov:2010ri} 
- has only recently been achieved within the framework of the AdS/CFT correspondence \cite{Benini:2015eyy,Cabo-Bizet:2018ehj,Choi:2018hmj,Zaffaroni:2019dhb,Cabo-Bizet:2017jsl}. On the macroscopic side, AdS black holes necessarily arise in gauged supergravity, where the presence of scalar potentials fixes a preferred symplectic frame, thereby obstructing a universal symplectically covariant treatment analogous to the asymptotically flat case. As a result, logarithmic corrections in AdS must be analyzed on a model-by-model basis.

Results obtained so far reveal a qualitative departure from flat-space black holes: the coefficient  $C$ in \eqref{Bekenstein+Log} appears to be generically non-topological, exhibiting explicit dependence on geometric data of the black hole background.  This behavior stands in tension with microscopic expectations, where logarithmic corrections are anticipated to be governed by topological or index-like quantities. Understanding the origin and interpretation of this non-universality in AdS, therefore, constitutes an important open problem.

A particularly sharp motivation for the present work comes from the black hole attractor mechanism
\cite{Ferrara_1995,Ferrara:1996dd,Ferrara:1996um,Strominger_1996}.
For BPS black holes in asymptotically flat spacetime, appearing in ungauged supergravity, the attractor mechanism is a classical statement asserting that scalar fields are driven to purely charge-dependent values at the horizon, independently of their asymptotic values \cite{Ferrara_1997}. In gauged supergravity, however, the presence of scalar potentials allows for flat directions in the attractor equations, and the statement for BPS black holes in AdS spacetime was refined as follows: 
the entropy is determined entirely by the charges and is independent of the values of the horizon moduli that are not fixed by the charges \cite{Cacciatori:2009iz}.
It is therefore natural to ask whether these classical flat directions persist once genuine quantum effects are taken into account.

In this work, we address this question by studying one-loop corrections to the entropy of static magnetically charged hyperbolic BPS black holes in asymptotically AdS$_4$ spacetime. These black holes arise as supersymmetric solutions of $\mathcal{N}=2$ Fayet–Iliopoulos (FI) gauged supergravity specified by the prepotential $F=-iX^0X^1$, and provide explicit examples that exhibit continuous families of horizon moduli 
at the classical level. For simplicity, we consider a bosonic consistent truncation of the gauged supergravity, which corresponds to an Einstein–Dilaton–Maxwell theory with a nontrivial scalar potential.

We compute the logarithmic corrections using heat kernel method \cite{Vassilevich:2003xt,Avramidi:1994fx,Hawking:1976ja,Denardo:1982mf,Barvinsky:2015bky}, taking into account both local contributions through Seeley-DeWitt coefficients and global contributions arising from zero modes. In particular, we find that the hyperbolic black hole horizon requires a different treatment of zero-mode counting compared to the case of compact horizons. The result of the zero mode counting is novel in the literature.  Our results show that the one-loop logarithmic correction  depends nontrivially on the unfixed horizon moduli $\nu$, leading to a  partition function of the form
\begin{eqnarray}\label{eq:Zresult}
    Z_{\text{AdS}_2} = e^{{\cal S}_{\mathrm{BH}}^{\text{class.}}}\int d\nu\; e^{\Delta {\cal S}(\nu)}\,, \qquad \Delta {\cal S}(\nu) = C(\nu)\ln \frac{A_H}{G_N}\,,
\end{eqnarray}
where the Bekenstein-Hawking entropy ${\cal S}_{BH}^{\text{class.}}$ is from the classical black hole near horizon saddle, and $\Delta{\cal S}(\nu)$ is from the one-loop correction around the saddle. We note that the path integral reduces to an ordinary integral over the moduli $\nu$. 
As a consequence, the logarithmic correction $\Delta{\cal S}(\nu)$ cannot be interpreted merely as a correction to the entropy, but instead plays the role of an effective quantum potential governing the integration over moduli in the quantum entropy function. We find that this quantum effect lifts the classical flat direction and stabilizes the moduli at the quantum level, providing an explicit example of a quantum realization of the attractor mechanism in gauged supergravity.

We also perform a number of nontrivial consistency checks of our one-loop computation. In particular, by considering various truncations of theory and the quadratic fluctuation operator, we verify that the general structure of the resulting Seeley-DeWitt coefficients correctly reproduces those of simpler theories,
 such as Einstein–Dilaton-Maxwell theories without scalar potentials, minimal gauged supergravity, and pure gravity theories. Applying these truncated results to the hyperbolic black hole backgrounds, we obtain the corresponding logarithmic corrections to the entropy,  which constitute additional results that have not been previously studied in the literature. 

It would be interesting to understand whether this moduli dependence persists in the fully supersymmetric theory, or whether additional cancellations occur once the complete supermultiplet spectrum is taken into account.
We leave a systematic analysis of the full supersymmetric theory, as well as a first-principles understanding of the integration measure over horizon moduli, for future work. 

The paper is organized as follows. In Section~\ref{HyperbolicBH}, we review the Einstein–Dilaton–Maxwell theory as a consistent truncation of ${\cal N}=2$ gauged supergravity and present the hyperbolic BPS black hole solutions. In Section~\ref{OneLoop}, we compute the one-loop corrections around the near-horizon geometry using heat kernel techniques, separating local and global contributions. Consistency checks of the one-loop computation are presented in Section~\ref{sec:Reduction}. We conclude in Section~\ref{sec:conclusion} with a discussion and outlook.
\section{Hyperbolic BPS black hole in AdS$_4$}
\label{HyperbolicBH}
\subsection{Truncation of $F = -i X^0 X^1$ gauged supergravity}
We consider a simple truncation of the ${\cal N}=2 $ FI-gauged supergravity coupled to a single vector multiplet specified by the prepotential $F= -i X^0 X^1$ (See the review on the gauged supergravity in appendix \ref{app:N=2SUGRA}). We consider the bosonic sector of the theory, and further truncation by considering only the real sector of the complex scalar $\tau \equiv X^1/X^0 $. The resulting theory is nothing but the Einstein-Dilaton-Maxwell theory, which consists of a real scalar, two Maxwell fields, and a graviton with a nontrivial scalar potential. With setting that {\color{black}{$8\pi G_N=1$}} 
, the Euclidean action is given by \footnote{This definition of Euclidean action is minus of conventional Euclidean action ${\cal S}= -S_E$ so that path integral is weighted by ${e}^{\cal S}$  as defined in \eqref{eq:QEF}. (The same convention was used in \cite{Sen:2012kpz}.)}
\footnote{From the action presented in \eqref{eq:lagrangian} following  \cite{Cacciatori:2009iz}, we have set $8\pi G_N=1$, and rescaled the gauge field  $A_\mu \rightarrow \sqrt{2} A_\mu$.}
\begin{equation}\label{eq:finalactiontau}
    {\cal S}
    ={\color{black}\frac{1}{2}} \int d^4x\,  \sqrt{ g\,} \left( R-\tau\, F^{(0)\mu\nu} F^{(0)}_{\mu\nu} 
    - \frac{1}{\tau}\, F^{(1)\mu\nu} F^{(1)}_{\mu\nu} 
    -    \frac{1}{2\tau^2}\,\partial_\mu \tau\, \partial^\mu \tau 
    - 2\,V(\tau)
    \right),
\end{equation}
where the potential is given by the Fayet-Ilopoulos gauging parameters $g_0$ and $g_1$ by
\begin{equation}\label{eq:scalarpottau}
    V(\tau)=     -\frac{2}{\tau}(g^2_0+4 g_0 g_1 \tau+g^2_1\tau^2)\,.
\end{equation}
Note that the range of the real scalar is $\tau >0 $ for positivity of the kinetic terms in the action, and thus, it will be natural to parametrize the scalar in terms of the dilaton as $\tau = e^{-2\Phi}$ when we consider the quadratic action.

The equations of motion are as follows:
for graviton, 
\begin{equation}\label{eq:eomg}
\begin{aligned}
0\;=\;& \; R_{\mu \nu} - \frac{1}{2}{g}_{\mu \nu} \left(R   - \tau{F}^{(0)}_{\,\rho\sigma} {F}^{(0)\rho\sigma}\,  
-\frac{1}{\tau}{F}^{(1)}_{\,\rho\sigma} {F}^{(1)\rho\sigma}\, - \frac{1}{2\tau^2} D_\rho \tau D^\rho \tau-2 V\, \right)\\
&\quad 
-  2  \tau {F}^{(0)}_{\,\mu \rho} {F}^{(0)}_{\,\nu}{}^\rho{} \, 
-\frac{2}{\tau}{F}^{(1)}_{\,\mu \rho} {F}^{(1)}_{\,\nu}{}^\rho{}  
- \frac{1}{2\tau^2} D_\mu \tau D_ \nu \tau \,,
\end{aligned}
\end{equation}
whose trace implies
\begin{equation}\label{eq:eomgTrace}
    R = 4 V\,,
\end{equation}
for the scalar field,
\begin{equation}\label{eq:eomtau}
\begin{aligned}
0= - D_\mu \left( \frac{1}{\tau}D^\mu \tau \right) + \tau    F^{(0)}_{\mu\nu}\, F^{(0)\mu\nu}
- \frac{1}{\tau}  F^{(1)}_{\,\mu\nu}\, F^{(1)\mu\nu} + 2\tau V'(\tau)\;,
\end{aligned}
\end{equation}
where  $2\tau V'(\tau) =   4 g_0^2\, \tau^{-1} - 4 g_1^2\, \tau $,  and for the Maxwell fields,
\begin{equation}\label{eq:eoma0}
0 =  D_\nu \left( \tau  F^{(0)\nu}{}_\mu \right)\,,\qquad 0 =  D_\nu \left( \frac{1}{\tau}  F^{(1)\nu}{}_\mu \right)\,.
\end{equation}
\subsection{Black hole near horizon background 
}\label{sec:BH background}
The theory admits a supersymmetric static black hole with magnetic charge and with a hyperbolic horizon \cite{Cacciatori:2009iz}. The full solution is reviewed in appendix \ref{app:AdS4blackhole}. Here, we consider the near-horizon geometry of the black hole solution as the horizon data that provides a self-consistent setting for the quantum entropy function. The geometry is given by AdS$_2 \times \mathbb{H}^2$,  
\begin{equation}\label{eq:NHmetric}
    ds^2= L_1^2 \bigg( \frac{d {r}^2}{{r}^2-1 }+({r}^2-1) d {\theta}^2\bigg)+L_2^2\left( d\psi^2 + \sinh^2\psi\, d\vartheta^2 \right)\;,
\end{equation}
where the radius of AdS$_2$ and the $\mathbb{H}^2$  respectively  are
\begin{equation}\label{eq:radius}
  L_1^2=(16  g_0 g_1 \cosh^2\nu )^{-1} \,,\qquad L_2^2=(8 g_0 g_1)^{-1}\;.
\end{equation}
This geometry is supported by the magnetic flux $p^I$ as
\begin{equation}
\begin{split}\label{eq:Fluxsolution}
F^I &
=\frac{1}{\sqrt{2}} 4\pi p^I d\mathbb{V}_{\mathbb{H}^2}
= \frac{1}{\sqrt{2}}4\pi p^I \sinh{\psi} d\psi \wedge d\vartheta\;,
\end{split}
\end{equation}
and the scalar field 
\begin{equation}\label{eq:tausolution}
    \tau = \frac{g_0}{g_1}e^{-2\nu}\,
\end{equation}
with unfixed real scalar parameter $\nu$.  
Maxwell equations together with the Bianchi identity constrain the magnetic charges $p^{I}$ and the gauging parameters $g_I$ to $ g_0p^0=g_1p^1$. This relation, along with Dirac
quantization condition, $8\pi g_I p^{I}=1$, enforces,
\begin{equation}
p^{I}=\frac{1}{16\pi g_I}\, .
\end{equation}

Note that the equations of motion do not completely fix the horizon value of the scalar field, leaving a continuous modulus $\nu$. Nevertheless,  the Bekenstein-Hawking entropy density is entirely determined by the quantized magnetic charge $p^I$ as
\begin{equation}\label{eq:entopy}
    \frac{{\cal S}_{\text{BH}}}{\mathbb{V}_{\mathbb{H}^2}} = \frac{1}{\mathbb{V}_{\mathbb{H}^2}}\frac{A_{\mathbb{H}^2}}{4}= \frac{L_2^2}{4}= 8\pi^2 p^0 p^1\,.
\end{equation}
This confirms the refined version of the classical attractor mechanism for gauged supergravity \cite{Cacciatori:2009iz}. The entropy formula can also be obtained by extremizing the entropy function~${\cal S}$ defined in \eqref{eq:QEF}. Since we consider a purely magnetic black hole, the electric charges vanish, $q_I =0$, and extremizing ${\cal S}$ is equivalent to the on-shell action~\eqref{eq:finalactiontau} on the near horizon solution. Reinstating the overall factor of $1/8\pi$ (with $G_N=1$) then reproduces the entropy density \eqref{eq:entopy}. Although the scalar potential is non-trivial, the $\nu$-dependence appearing in $V'(\tau)$ on the near-horizon background is canceled by its coupling to the gauge field strengths, as follows from the scalar equation of motion~\eqref{eq:scalareom}.
We also note that the volume of the hyperbolic space $\mathbb{V}_{\mathbb{H}^2}$ is infinite, so we work with the entropy density. 
\footnote{
While the hyperbolic plane $\mathbb{H}^2$ is locally equivalent to Euclidean AdS$_2$, its volume in the black-hole entropy should be understood either as a density or as the volume of a compact quotient, rather than as the holographically regularized value $-2\pi$. }

We further note that in the limit $\nu \rightarrow 0$, the hyperbolic BPS black hole solution reduces to a solution of minimal gauged supergravity discovered in \cite{Caldarelli:1998hg}.
We also note that there is no ungauged supergravity limit:  it is clear from the \eqref{eq:radius} that $g_0, g_1 \rightarrow 0$ implies the zero radius of the solution. Also, at the level of the theory, the potential \eqref{eq:scalarpottau} vanishes, and thus the cosmological constant is absent, which implies that the black holes in AdS$_4$ spaces are not admitted.

For later use, we shall present the background invariants as follows:
\begin{align}\label{eq:backgroundinvr}
    &R = -2\left(\frac{1}{L^2_1}+\frac{1}{L^2_2}\right)= -16 g_0 g_1 (2+ \cosh 2\nu)\,\,\,,\,\,\quad \nonumber
    \\
    &R_{\mu\nu}R^{\mu\nu} =\frac{1}{2}R_{\mu\nu \rho\sigma}R^{\mu\nu \rho\sigma} =2\left(\frac{1}{L^4_1}+\frac{1}{L^4_2}\right) = 64 g^2_0 g^2_1 \left(5+4\cosh{2\nu}+\cosh{4\nu}\right)\,,
   \\
&     e^{-2\phi_0} (F^{(0)})^2 = 4g_0 g_1 e^{-2\nu}\,,\qquad  e^{2\phi_0}(F^{(1)})^2  = 4g_0 g_1 e^{2\nu} \,.\nonumber
\end{align}

\section{One-loop correction}\label{OneLoop}
In this section, we compute the one-loop corrections to the entropy function $\Delta {\cal S}(\nu)$ as described in \eqref{eq:Zresult}. As the main technical tool, we employ the heat kernel expansion method \cite{Vassilevich:2003xt,Avramidi:1994fx,Hawking:1976ja,Denardo:1982mf,Barvinsky:2015bky}. We begin by reviewing the heat kernel setup, then present the explicit form of the relevant quadratic fluctuation operators for bosonic and ghost fields and the associated local heat kernel coefficients, and finally extract the logarithmic corrections, including both local and global (zero-mode) contributions.
\subsection{Heat kernel setup}
Let us split the fields $\Psi$ into their background value and the fluctuations as
\begin{equation}
    \Psi = \bar{\Psi}(\nu) + \delta \Psi\,.
\end{equation}
Here, we consider the background field $\bar{\Psi}$ as the black hole near-horizon configuration presented in the previous section, which depends on the unfixed moduli parameter $\nu$. Then the one-loop correction to the entropy function is obtained by the functional integration as 
\begin{equation}
    \Delta {\cal S} = \ln Z_{\text{1-loop}}\,, \qquad Z_{\text{1-loop}}=\int D[ \delta \Psi] \exp\!\left({\cal S}^{(2)}[\delta \Psi]
    \right)\,,    
\end{equation}
where~${\cal S}^{(2)}[\delta \Psi]$ is the quadratic action in the fluctuations.
Since the fluctuation $\delta\Psi$ includes the zero mode of the quadratic action, we further split them into the zero mode and the non-zero mode fluctuations
\begin{equation}
    \delta \Psi = \delta \Psi_{\text{zm}} + \xi\,. 
\end{equation}
Then, the non-trivial part of the quadratic action is schematically written as 
\begin{equation}\label{S(2)}
{\cal S}^{(2)} \;=\; \frac{1}{2}
 \int d^4 x \, \sqrt{\bar{g} } \;\,
\xi_m \, {\cal H}^{mn}\,   \xi_n \,.
\end{equation}
The one-loop partition function splits into two contributions from non-zero mode Gaussian integration, and the zero mode path integral measures as
\begin{equation}\label{eq:split_lnZ}
   \ln Z_{\text{1-loop}} =\ln  Z_{\text{1-loop}}^{\text{nzm}}+ \ln  Z_{\text{1-loop}}^{\text{zm}} \,.
\end{equation}
The total contribution to the correction of entropy can be summarized as the sum of the local and global parts as
\begin{equation}\label{eq:DeltaS}
    \Delta {\cal S}
    = \frac{1}{2}\Bigl( C_{\text{local}}+C_{\text{global}}\Bigr)
    \ln \frac{A_H}{G_N}\, .
\end{equation}
where the local and global part, $C_{\text{local}}$ and $C_{\text{global}}$ are defined as 
\begin{eqnarray}
    \label{eq:Clocal}
    C_{\text{local}}&\equiv& \int d^4 x \sqrt{g}\;a_4(x)\,,
\qquad 
    C_{\text{global}}\;\equiv\; \bigl(\beta_\Psi-1 \bigr)n_\Psi^{\text{zm} } \,.
\end{eqnarray}
which shall be explained in the following way. 

The $C_{\text{local}}$ and  the $- n^{\text{zm}}_{\Psi}$ part of the $C_{\text{global}}$ in \eqref{eq:Clocal} are originating from the non-zero mode contribution $\ln Z^{\text{nzm}}_{\text{1-loop}}$ in \eqref{eq:split_lnZ} as follows: \begin{align}\label{eq:lnZ_1-loop}
    \begin{split}
         \ln Z_{\text{1-loop}}^{\text{nzm}} &= -\frac{1}{2}\ln {\det}' {\cal H} = -\frac{1}{2}{\sum_{n}}^\prime \,\ln \frac{\kappa_n}{L^2}
         \\
         &= \frac{1}{2}\lim_{\epsilon \rightarrow 0}\int_{\epsilon}^{\infty} \frac{ds}{s}{\sum_{n}}^{\prime}e^{-s \frac{\kappa_n}{L^2}} = \frac{1}{2}\lim_{\epsilon \rightarrow 0}\int_{\epsilon/L^2}^{\infty} \frac{d\bar{s}}{\bar{s}}{\sum_{n}}^{\prime}e^{-\bar{s}\kappa_n}
         \\
         &= \frac{1}{2}\lim_{\epsilon \rightarrow 0}\int_{\epsilon/L^2}^{\infty} \frac{d\bar{s}}{\bar{s}} \left(K(\bar{s})- n^{\text{zm}}_\Psi\right)\,.
    \end{split}
\end{align}
Here,  $\sum^\prime$ denotes the sum over non-zero modes and $\kappa_n/L^{2} $ denotes the eigenvalue of the operator ${\cal H}$ with dimensionless $\kappa_n$ and the length scale parameter $L$. To the second line, we use the integral representation of the logarithm by introducing the Schwinger proper-time parameter $s$  and the UV cut-off $\epsilon$, and then rescale the parameter to the dimensionless parameter $\bar{s}$. To the third line, we add and subtract the number of zero modes $n^{\text{zm}}_{\Psi}$ to make the summation over a complete functional basis and define the trace of the heat kernel,
\begin{equation}\label{eq:heatkernel}
    K(\bar{s}) \equiv \sum_n e^{-\bar{s}\kappa_n}= \int d^dx\sqrt{g} \sum_n e^{-\bar{s}\kappa_n}\Psi^\ast_n(x)\Psi_n(x) \equiv \int d^dx\sqrt{g} K(x,\bar{s})\,.
\end{equation}
We can then use the heat kernel expansion with respect to the small $\bar{s}$ parameter as
\begin{equation}\label{eq:seeleydewitt}
    K(x,\bar{s})= \sum_{n\geq0}\bar{s}^{n-{2}}a_{2n}(x)= \bar{s}^{-{2}}a_0(x)+ \bar{s}^{-1}a_2(x)+ a_{4}(x) + {\cal O}(\bar{s})\,,
\end{equation}
where the expansion coefficients $a_{2n}(x)$ are known as Seeley-DeWitt coefficients. The first two terms in the expansion provide the power-law divergences in the integral \eqref{eq:lnZ_1-loop}, where the coefficients $a_0$ and $a_2$ respectively correspond to one-loop renormalisations of cosmological constant and that of Newton's constant and electric charge \cite{David:2021eoq}. Since our interest is in the logarithmic correction, we focus on the $\bar{s}$ independent term, $a_4(x)$. Therefore, the integration in the last line of \eqref{eq:heatkernel} gives the $C_{\text{local}}$ and the $-n^{\text{zm}}_\Psi$ part of $C_{\text{global}}$ as defined in \eqref{eq:Clocal}. 

The remaining part $-\beta_\Psi n^{\text{zm}}_\Psi$ of $C_{\text{global}}$ in \eqref{eq:Clocal} is from zero modes contribution in $\ln Z^{\text{zm}}_{\text{1-loop}}$. The zero modes are associated with the pure gauge modes of the gauge fields with non-normalizable parameters, yet the modes are normalizable.  Since the parameter is not normalizable, they should be considered as  physical degrees of freedom. Since they are zero modes of the theory, the contribution is only through the integration measure.    Although the integration diverges, we can extract the length scale $L$ dependence as
\begin{equation}
    Z_{\text{1-loop}}^{\text{zm}} = \int D[\delta\Psi_{\text{zm}}] \sim L^{\beta_\Psi n_\Psi^{\text{zm}}}\,.
\end{equation}
The length scale is related to the horizon area $L^2 \sim A_{H}$. 
The power $\beta_{\Psi}$ to the length scale~$L$ is determined by requiring ultra-local argument, i.e. demanding the normalization condition by $1 =\int D[\delta \Psi_{\text{zm}}]\exp \left( -||\delta \Psi_{\text{zm}}||^2\right)$. In four-dimensions, the $\beta$ for graviton and the one-form gauge field are summarized by \cite{Sen:2012cj} 
\begin{equation}\label{eq:beta}
    \beta_g = 2\,,\qquad \beta_{\text{1-form}}=1\,.
\end{equation}
Therefore, from the formula in \eqref{eq:Clocal}, one can see that only the graviton zero modes contribute to the $C_{\text{global}}$. The number of zero modes will be counted in the later section~\ref{sec:ZM}.
\subsubsection*{Seeley DeWitt coefficient}
Once the quadratic operator is identified, then the Seeley DeWitt coefficient $a_4(x)$ can be obtained as follows. 

The most general second-order operator ${\cal H}_{mn}$ arising from the quadratic action \eqref{S(2)} can be aligned with the following canonical structure
\begin{equation}\label{eq:kineticop_short}
    \xi_m {\cal H}^{mn}{\xi}_n
    = \,{\xi}_m\Big[(D_\rho D^\rho) I^{mn} + N_\rho^{mn}D^\rho + P^{mn}\Big]{\xi}_n\;,
\end{equation}
where $D_\rho$ is the background covariant derivative, $I^{mn}$ is the effective identity (or metric) in field space, $N_\rho^{mn}$ encodes first-derivative couplings, and $P^{mn}$ is the non-derivative term in the kinetic operator.

To obtain the minimal Laplace-type form required for the heat-kernel method, we absorb all terms linear in derivatives into an effective connection and define the remaining non-derivative matrix-valued potential term as $E$,
\begin{equation}\label{eq:connection_short}
    \omega_\rho^{mn}\equiv \tfrac{1}{2}N_\rho^{mn},\qquad
    E^{mn}=P^{mn}-(D^\rho\omega_\rho)^{mn}-(\omega_\rho)^{mp}(\omega^\rho)_p{}^{\,n}.
\end{equation}
In this form, the kinetic operator reads compactly as
\begin{eqnarray}
{\cal H}=\mathcal{D}_\rho \mathcal{D}^\rho\,I + E\;,
\end{eqnarray}
with $\mathcal{D}_\rho$ now understood to include $\omega_\rho$, i.e. ${\cal D}_\rho = D_\rho + \omega_\rho$.
Then, the curvature of this connection acting on the fluctuation bundle is 
\begin{equation}\label{eq:Omega_short}
    (\Omega_{\rho\sigma})^{mn}\equiv [{\cal D}_\rho\,,{\cal D}_\sigma]=[D_\rho,D_\sigma]^{mn}+{\color{black}2}D_{[\rho}\omega_{\sigma]}^{mn}+[\omega_\rho,\omega_\sigma]^{mn}.
\end{equation}

With  the above canonical building blocks of the quadratic operator, the Seeley-DeWitt coefficient $a_4$ in the heat-kernel expansion  \eqref{eq:seeleydewitt} is obtained as \cite{Vassilevich:2003xt}:
\begin{equation}\label{eq:seeley_short}
\begin{aligned}
    a_4(x) &= \frac{\chi}{16\pi^2\times 360}\,\mathrm{Tr}\Big(60RE+180E^2+30\Omega_{\rho\sigma}\Omega^{\rho\sigma}\\[-2pt]
    &\qquad\qquad\qquad\qquad\qquad\qquad+(2R_{\mu\nu\rho\sigma}R^{\mu\nu\rho\sigma}-2R_{\mu\nu}R^{\mu\nu}+5R^2)\,I\Big)\,,
\end{aligned}
\end{equation}
where the overal sign factor $\chi=1$ for bosonic fields and $\chi=-1$ for Grassmann odd fields.
Here, we note that the effective identity $I^{mn}$ appearing in~\eqref{eq:kineticop_short} serves as a metric in the field space. The index $m$ is raised and lowered by the metric $I^{mn}$ and $I_{mn}$, respectively. (see Appendix~\ref{app:detailedcal} for the definition and examples.)

In the next subsection, we will explicitly obtain the second-order operator to read off the $I^{mn}$, $\omega_\rho^{mn}$, $P^{mn}$, and then $\Omega^{mn}$ and $E$ to obtain the Seeley-DeWitt coefficient \eqref{eq:seeley_short}. The curvature invariants $R$, $R_{\mu\nu}R^{\mu\nu}$ and $R_{\mu\nu\rho\sigma}R^{\mu\nu\rho\sigma}$ are given in \eqref{eq:backgroundinvr}.
\subsection{Quadratic action to Seeley DeWitt coefficient}
\subsubsection{From gauged fixed action}\label{sec:gaugedfixed}

To find the quadratic action, let us first parametrize the scalar as 
\begin{equation}
    \tau = e^{-2\Phi}\,,
\end{equation}
to manifest the information $\tau >0$ and  consider the action in the form of 
\begin{equation}\label{eq:Action_Phi}
   {\cal S}=\frac{1}{2}\int d^4 x \sqrt { g}\left(R-e^{-2\Phi} F^{(0)\mu\nu}F^{(0)}_{\mu\nu}-e^{2\Phi}F^{(1)\mu\nu}F^{(1)}_{\mu\nu}-2\partial_\mu\Phi \partial^\mu \Phi-2 V(\Phi)\right)\;,
\end{equation}
where the potential is written as  $ V(\Phi)= -8 g_0 g_1 - 2g_0^2\,e^{2\Phi}  -2 g_1^2\,e^{-2\Phi}$.

From this action, we find the quadratic terms of the quantum fluctuation around the classical saddle. Denoting the classical background by,
$\phi_0\,,\bar{g}_{\mu\nu}\,, \bar{A}^{(I)}_\mu$, we consider the quantum fluctuations 
\begin{equation}\label{eq:fluctuations}
     \Phi = \phi_0 + \frac{1}{\sqrt{2}}\phi\,,\quad g_{\mu\nu}=\bar{g}_{\mu\nu}+2\,h_{\mu\nu}, 
    \quad 
    A^{(0)}_\mu=\bar{A}^{(0)}_\mu+ \frac{1}{\sqrt{2}}e^{\phi_0}a^{(0)}_\mu \,,
    \quad 
    A^{(1)}_\mu=\bar{A}^{(1)}_\mu+ \frac{1}{\sqrt{2}}e^{-\phi_0}a^{(1)}_\mu\\,
\end{equation}
where we have used the appropriate renormalization for the fluctuations with numerical factors and background value of scalar $\phi_0$ in order to obtain a canonically normalized Laplacian operator in the quadratic action. 

 Together with the action \eqref{eq:Action_Phi}, we also need to consider gauge fixing terms as well as ghost actions. For the diffeomorphism and $U(1)^2$ gauge symmetries, we impose the harmonic gauge and the Lorentz gauge that correspond to adding the gauge fixing terms, respectively, as
 \begin{align}
     \begin{split}
        \label{eq:gaugefix1}
 S^{h}_{\text{g.f.}} &= - \int d^4x \, \sqrt{ g} \, 
\left( D^\nu h_{\nu\rho} - \frac{1}{2} D_\rho h^\beta_{\ \beta} \right)
\left( D^\mu h_{\mu}^{\ \rho} - \frac{1}{2} D^\rho h^\alpha_{\ \alpha} \right),
\\
  S^{a}_{\text{g.f.}} &= -\frac{1}{2} \int d^4x \, \sqrt{ {g}} \, \left((D_\mu a^{(0)\mu})^2+(D_\mu a^{(1)\mu})^2\right)  \,.  
     \end{split}
 \end{align}
 Since we will consider quadratic action, the measure $\sqrt{g}$ and the covariant derivatives are in their background values.   We also need to add the ghost fields actions, which will be considered in the next subsection \ref{sec:ghosts}.

Now, we proceed to obtain the quadratic terms from the gauge-fixed action, i.e., the physical action \eqref{eq:Action_Phi} together with the gauge-fixing terms \eqref{eq:gaugefix1}, with respect to the field fluctuations in \eqref{eq:fluctuations}. We organize the fluctuations of the physical fields as
\begin{equation}
     {\xi}_m = \{ h_{\mu\nu},\, a^{(0)}_\rho,\, a^{(1)}_\sigma,\, \phi\}\,.
\end{equation} 
 Since the action contains several terms, it is convenient to analyze the quadratic action term by term. Let us consider the total gauged fixed action as 
 \begin{equation}
     {\cal S}_{\text{gauge-fixed}}= \int d^4 x \left({\cal L}_{h} + {\cal L}_{a_{(0)}}+ {\cal L}_{a_{(1)}}+ {\cal L}_{\phi}\right)\,,
 \end{equation}
 where each Lagrangian density denotes the Einstein-Hilbert term,  the two gauge field kinetic terms, 
 and the scalar terms with its potential in \eqref{eq:Action_Phi}, with addition of the corresponding gauge fixing terms in \eqref{eq:gaugefix1}. Then, we obtain the quadratic Lagrangian density as follows. 
 
From the Einstein-Hilbert term with gauge fixing, we obtain
\begin{equation}\label{eq:EHvariation}
\begin{split}
  {\cal L}^{(2)}_{h}& = \frac{1}{2} \sqrt{ \bar{g}} \, \Big[ 
h^{\mu\nu} D^\rho D_\rho h_{\mu\nu} 
- \frac{1}{2} h^\mu_{\ \mu} D^\rho D_\rho h^\nu_{\ \nu} + 2 h^{\mu\nu} h^{\rho\sigma}\bar R_{\mu\rho\nu\sigma}
\\& 
\qquad\qquad {\color{black}
+2h^{\mu\nu}\bar{R}_{\mu\rho}h^\rho{}_{\nu}
 }
- 2 h^{\mu\nu} h^\rho_{\ \rho} \bar R_{\mu\nu}- \left( h^{\alpha\beta} h_{\alpha\beta} 
- \frac{1}{2} h^\alpha_{\ \alpha} h^\beta_{\ \beta} \right)\bar R
\Big]\,.
\end{split}
\end{equation} 
 From the kinetic term of gauge field $a^{(0)}$ with gauge fixing, we obtain
\begin{eqnarray}\label{eq:a0qudratic}
    {\cal L}^{(2)}_{a_{(0)}}&=& \frac{1}{2}\sqrt{\bar{g}} \left[ a^{(0)}_\mu  D_\rho D^\rho a^{(0)\mu}-a^{(0)}_{\mu}\bar R^{\mu\nu}a^{(0)\nu } - \phi^2  e^{-2\phi_0} (\bar{F}^{(0)})^2  \right.\nonumber
    \\
    && \qquad\quad  +  h_{\mu\nu}h_{\lambda \sigma }\, e^{-2\phi_0} \left\{  \left(  \bar{g}^{\mu\lambda}\bar{g}^{\nu\sigma}-\tfrac{1}{2}\,\bar{g}^{\mu\nu}\bar{g}^{\lambda \sigma}  \right)(\bar{F}^{(0)})^2 - 4\bar{F}^{(0)\mu\lambda}\bar{F}^{(0)\nu\sigma} \right. \nonumber
    \\
    &&\qquad \qquad \qquad \quad \qquad \quad \left. +\, 4\, \bar{g}^{\mu\nu}\bar{F}^{(0)\lambda}{}_{\kappa}\bar{F}^{(0)\sigma \kappa } - 8\, \bar{F}^{(0)\mu }{}_{\kappa}\bar{F}^{(0)\lambda \kappa}\bar{g}^{\nu\sigma } \right\}
     \\
    &&\qquad\quad   + \sqrt{2}\phi \,h_{\mu\nu}\,e^{-2\phi_0}\left(\,\bar{g}^{\mu\nu}(\bar{F}^{(0)})^2-4\bar{F}^{(0)\mu}{}_{\rho}\bar{F}^{(0)\nu\rho}  \right)\nonumber
    \\
    &&\qquad\quad +  
       2\sqrt{2} h_{\mu\nu}\,e^{-\phi_0} \left( 2 \bar{g}^{\mu \lambda }\bar{F}^{(0)\nu\rho}- 2\bar{g}^{\mu\rho}\bar{F}^{(0)\nu\lambda} - \bar{g}^{\mu\nu} \bar{F}^{(0)\lambda\rho}\right)D_{\lambda}a^{(0)}_{\rho}\nonumber 
       \\
       &&\qquad \quad \left.+ 4\,\phi \,e^{-\phi_0}\bar{F}^{(0)\lambda \rho}D_{\lambda}a^{(0)}_{\rho} \right]\,.\nonumber
\end{eqnarray}
The term for the gauge field $a^{(1)}$ can easily be obtained by replacing $\{\phi_0,\, \phi\} \rightarrow \{-\phi_0,\,-\phi\}$ as
\begin{eqnarray}\label{eq:a1qudratic}
    {\cal L}^{(2)}_{a_{(1)}}&=& \frac{1}{2} \sqrt{\bar{g}} \left[ a^{(1)}_\mu  D_\rho D^\rho a^{(1)\mu}-a^{(1)}_{\mu}\bar R^{\mu\nu}a^{(1)\nu } - \phi^2  e^{2\phi_0} (\bar{F}^{(1)})^2  \right.\nonumber
    \\
    &&\qquad \quad +  h_{\mu\nu}h_{\lambda \sigma }\, e^{2\phi_0} \left\{  \left(  \bar{g}^{\mu\lambda}\bar{g}^{\nu\sigma}-\tfrac{1}{2}\,\bar{g}^{\mu\nu}\bar{g}^{\lambda \sigma}  \right)(\bar{F}^{(1)})^2 - 4\bar{F}^{(1)\mu\lambda}\bar{F}^{(1)\nu\sigma} \right. \nonumber
    \\
    &&\qquad \qquad \qquad \quad  \qquad  \quad\left. +\, 4\, \bar{g}^{\mu\nu}\bar{F}^{(1)\lambda}{}_{\kappa}\bar{F}^{(1)\sigma \kappa } - 8\, \bar{F}^{(1)\mu }{}_{\kappa}\bar{F}^{(1)\lambda \kappa}\bar{g}^{\nu\sigma } \right\}
     \\
    &&\qquad \quad  + \sqrt{2}\phi \,h_{\mu\nu}\,e^{2\phi_0}\left(4\bar{F}^{(1)\mu}{}_{\rho}\bar{F}^{(1)\nu\rho}-\,\bar{g}^{\mu\nu}(\bar{F}^{(1)})^2  \right)\nonumber
    \\
    &&\qquad \quad +  
       2\sqrt{2} h_{\mu\nu}\,e^{\phi_0} \left( 2 \bar{g}^{\mu \lambda }\bar{F}^{(1)\nu\rho}- 2\bar{g}^{\mu\rho}\bar{F}^{(1)\nu\lambda} - \bar{g}^{\mu\nu} \bar{F}^{(1)\lambda\rho}\right)D_{\lambda}a^{(1)}_{\rho}\nonumber 
       \\
       &&\qquad \quad \left.- 4\,\phi \,e^{\phi_0}\bar{F}^{(1)\lambda \rho}D_{\lambda}a^{(1)}_{\rho} \right]\,.\nonumber
\end{eqnarray}
Finally, from the terms for the scalar field, we obtain
\begin{eqnarray}\label{eq:scalarquadratic}
    {\cal L}^{(2)}_{\phi} &=&\frac{1}{2}  \sqrt{\bar{g}}\left[ h_{\mu\nu}h_{\lambda \sigma }(\bar{g}^{\mu \lambda}\bar{g}^{\nu \sigma } + \bar{g}^{\lambda\mu}\bar{g}^{\sigma \nu}- \bar{g}^{\mu\nu}\bar{g}^{\lambda \sigma})V(\phi_0)\right. 
  \\
  &&\qquad\quad \left.  -\partial_\mu \phi\partial^\mu \phi -\frac{1}{2}\phi^2 V''(\phi_0)-\sqrt{2}\, \phi\, h_{\mu\nu}\bar{g}^{\mu\nu }  V'(\phi_0)  \right]\,,\nonumber
\end{eqnarray}
where the background values associated with the potential are given by
\begin{equation}
    V(\phi_0) = - 4g_0 g_1(2 + \cosh 2\nu)  \,, V'(\phi_0) = -8 g_0 g_1 \sinh 2\nu \,, V''(\phi_0) = -16 g_0 g_1 \cosh 2\nu\,.
\end{equation}
\subsection*{Identifying the quadratic operator}
Collecting all the quadratic Lagrangian densities in \eqref{eq:EHvariation}, \eqref{eq:a0qudratic}, \eqref{eq:a1qudratic}, and \eqref{eq:scalarquadratic}, we identify the effective metric $I^{mn}$,  the gauge connection $N_\rho^{mn}\equiv 2(\omega_\rho)^{mn} $, and the mass matrix $P^{mn}$ of the canonical form of the quadratic operator~\eqref{eq:kineticop_short}. 

To obtain the simplified expression, we  use the equations of motion and the properties of the background fields: since the scalar background is constant, the scalar equation of motion \eqref{eq:eomtau} becomes
    \begin{align}
\begin{split}\label{eq:scalareom}
  &   V'(\phi_0)- e^{-2\phi_0}(\bar{F}^{(0)})^2+ e^{2\phi_0}(\bar{F}^{(1)})^2 
  =0\,,
  \end{split}        
    \end{align}
    which replaces the $V'(\phi_0)$ by the square of the field strengths. 
    The graviton equation of motion is followed by
   \begin{align}
\begin{split}\label{eq:gravitoneom}
e^{-2\phi_0}\bar{F}^{(0)}_{\mu\rho}\bar{F}^{(0)}_{\;\; \nu}{}^{\rho} + e^{2\phi_0}\bar{F}^{(1)}_{\mu\rho}\bar{F}^{(1)}_{\;\; \nu}{}^{\rho} 
    = \tfrac{1}{2}R_{\mu\nu }  -\tfrac{1}{4}g_{\mu\nu} \left(\tfrac{1}{2}R - e^{-2\phi_0}(\bar{F}^{(0)})^2- e^{2\phi_0}(\bar{F}^{(1)})^2 \right)\,.
\end{split}        
    \end{align}
   The trace of the graviton equation of motion
    \begin{equation}
        4V_0 = R\,,
    \end{equation}
    replaces the potential by the curvature. 
    Furthermore, we use the result of the solution that the effective mass of the scalar field vanishes:
    \begin{equation}\label{eq:no_Mass}
        \frac{1}{2}V''(\phi_0)+ e^{-2\phi_0}(\bar{F}^{(0)})^2+ e^{2\phi_0}(\bar{F}^{(1)})^2 =0\,.
    \end{equation}
Maxwell equations of motion and the Bianchi identity give
   \begin{equation}\label{eq:bianchi}
       D_\mu \bar F^{(I)\mu\nu}=0\,,\qquad  D_{[\mu}\bar F^{(I)}_{\nu\rho]} =0\,.
   \end{equation}
Additional useful identities of background fields based on the above equations of motion are listed in \eqref{eq:rule1} to \eqref{ruleF2=R2}. 
   
Furthermore, we manifest the symmetric properties in their expressions. Since   the second-order operators for the bosonic real fluctuations $\xi_m$ are symmetric, we use the following symmetrization relations: 
\begin{equation}\label{eq:symmetricprop}
\begin{split}
    &\xi_m \mathbf{A}^{mn} \xi_n =\frac{1}{2}\xi_m \mathbf{A}^{mn}\xi_n + \frac{1}{2}\xi_n \mathbf{A}^{mn} \xi_m \,,
    \\
    & \xi_m \mathbf{(B^\rho)}^{mn}D_\rho \xi_m=
 \frac{1}{2} \xi_m   \mathbf{(B^\rho)}^{mn}D_\rho \xi_n -\frac{1}{2}\,\xi_n \,\mathbf{(B^\rho)}^{mn} D_\rho \xi_m \\&\qquad \qquad \qquad \qquad -\frac{1}{4} \xi_m (D_\rho\mathbf{B^\rho)}^{mn}\xi_n - \frac{1}{4}\xi_n (D_\rho\mathbf{B^\rho)}^{mn} \xi_m\,. 
\end{split}
\end{equation}

The effective metric $I^{mn}$ for the present set of off-shell fluctuations $\{\xi_m\}$ can be read from the quadratic form as,
\begin{equation}\label{eq:Imn}
    \begin{split}
        \xi_m I^{mn} D_\rho D^\rho \xi_n
        &= \; h_{\mu\nu}\;I^{\mu\nu, \lambda\sigma} D_\rho D^\rho h_{\lambda \sigma}
       \\ 
        & \;\;+a^{(0)}_{\mu}\bar g^{\mu\nu} D_\rho D^\rho a^{(0)}_{\nu}+a^{(1)}_{\mu} \bar g^{\mu\nu} D_\rho D^\rho a^{(1)}_{\nu}+ \phi\, D_\rho D^\rho \phi\,.
    \end{split}
\end{equation}
Here, the matrix $I^{\mu\nu,\lambda\sigma}$ is called the DeWitt metric, defined by (see Appendix \ref{app:Identities} for more details.)
\begin{equation}
   I^{\mu\nu,\lambda\sigma} = \frac{1}{2}\bigl(\bar{g}^{\mu\lambda}\bar{g}^{\nu\sigma}+\bar{g}^{\mu\sigma}\bar{g}^{\nu\lambda}- \bar{g}^{\mu\nu}\bar{g}^{\lambda\sigma}\bigr)\,,
\end{equation}
whose inverse is $I_{\mu\nu,\lambda\sigma} = \frac{1}{2}\bigl(\bar{g}_{\mu\lambda}\bar{g}_{\nu\sigma}+\bar{g}_{\mu\sigma}\bar{g}_{\nu\lambda}- \bar{g}_{\mu\nu}\bar{g}_{\lambda\sigma}\bigr)$  
such that $ I^{\mu\nu,\lambda\sigma}I_{\lambda \sigma, \delta\rho} = \delta^{(\mu}_{\rho}\delta^{\nu)}_{\delta}$.
Therefore, the full metric $I^{mn}$ is the $19\times 19$ matrix,
\begin{equation}
   I^{mn}=  \mathrm{diag} \Bigl\{ \frac{1}{2}\bigl(\bar{g}^{\mu\lambda}\bar{g}^{\nu\sigma}+\bar{g}^{\mu\sigma}\bar{g}^{\nu\lambda}- \bar{g}^{\mu\nu}\bar{g}^{\lambda\sigma}\bigr)\,, \bar{g}^{\delta\rho}\,,\bar{g}^{\kappa\eta}\,, 1 \Bigr\}\,,
\end{equation}
that acts on the field space $(h_{\lambda\sigma}\,, a^{(0)}_\rho \,,a^{(1)}_{\eta}\,,\phi )$.\\
From the derivative-independent part of the operator, one can identify the matrix $P$ and determine its non-zero components as shown below.
\begin{eqnarray}\label{eq:Pmn}
    \xi_m P^{mn}\xi_n 
    &=&h_{\mu\nu}  \Big\{
    I^{\mu\nu,\lambda\sigma}\left(\tfrac{1}{2}\bar R -e^{-2\phi_0} (\bar F^{(0)})^2-e^{2\phi_0}  (\bar F^{(1)})^2 \,\right)\nonumber
    \\
&&\qquad \; +\,4\,e^{-2\phi_0}\bar F^{(0)\mu(\lambda}  \bar F^{(0)\sigma)\nu}+4\,e^{2\phi_0} \bar F^{(1)\mu(\lambda} \bar F^{(1)\sigma)\nu}\nonumber
\\
&& \qquad \; - 2 \bar R^{\mu(\lambda \sigma )\nu  } \textcolor{black}{- \bar  R^{\mu(\lambda }\bar{g}^{\sigma) \nu} 
-
\bar R^{\nu(\lambda }\bar{g}^{\sigma) \mu}
} \Big\}h_{\lambda\sigma}
\nonumber
\\
&& -a^{(0)}_\mu  \bar R^{\mu\nu}a^{(0)}_\nu-a^{(1)}_\mu \bar R^{\mu\nu} a^{(1)}_\nu \nonumber
\\
&& - \sqrt{2}\, h_{\mu\nu}D^{(\mu }\left(e^{-\phi_0} \bar{F}^{(0)\nu)\sigma}\right)a^{(0)}_{\sigma}
- \sqrt{2}\, h_{\mu\nu}D^{(\mu }\left(e^{\phi_0} \bar{F}^{(1)\nu)\sigma}\right)a^{(1)}_{\sigma}\nonumber
\\
&&
- \sqrt{2}\,a^{(0)}_{\sigma} D^{(\mu }\left(e^{-\phi_0} \bar{F}^{(0)\nu)\sigma}\right)h_{\mu\nu}
- \sqrt{2}\,a^{(1)}_{\sigma} D^{(\mu }\left(e^{\phi_0} \bar{F}^{(1)\nu)\sigma}\right)h_{\mu\nu}\,\nonumber
  \\
&& + 2\sqrt{2} \, h_{\mu\nu} \Bigl( - e^{-2\phi_0 }\bar{F}^{(0)\mu}{}_{\rho}\bar{F}^{(0)\nu \rho} +e^{2\phi_0 }\bar{F}^{(1)\mu}{}_{\rho}\bar{F}^{(1)\nu \rho}\Bigr)\phi 
\\
&& + 2\sqrt{2} \,\phi\,\Bigl( - e^{-2\phi_0 }\bar{F}^{(0)\mu}{}_{\rho}\bar{F}^{(0)\nu \rho} +e^{2\phi_0 }\bar{F}^{(1)\mu}{}_{\rho}\bar{F}^{(1)\nu \rho}\Bigr) h_{\mu\nu} \nonumber\;.
\end{eqnarray}
Here we note that the effective mass term of $\phi$ vanishes due to the relation \eqref{eq:no_Mass}. The first derivative terms come from the symmetrization process \eqref{eq:symmetricprop} using the integration by parts.
In addition, the gauge-connection matrix $\omega_\rho$ is obtained by analysing the linear-derivative contributions from each term individually and then combining them in the operator form. It is antisymmetric with respect to the associated fluctuations and is given by
   \begin{align}\label{eq:omega}
       \begin{split}
       2\xi_m (\omega^\rho)^{mn}D_\rho \xi_n =&\;\,
        \sqrt{2}\,h_{\mu\nu}\, e^{-\phi_0}\left(2\bar{F}^{(0)\rho (\mu  }\bar{g}^{\nu) \sigma }-2\bar{F}^{(0)\sigma (\mu  }\bar{g}^{\nu) \rho } -\bar{g}^{\mu\nu}\bar{F}^{(0)\rho \sigma}\right)D_{\rho}\,a^{(0)}_{\sigma} 
       \\
       & +\sqrt{2}\,h_{\mu\nu} \,e^{\phi_0}\left(2\bar{F}^{(1)\rho (\mu  }\bar{g}^{\nu) \sigma }-2\bar{F}^{(1)\sigma (\mu  }\bar{g}^{\nu) \rho } - \bar{g}^{\mu\nu }\bar{F}^{(1)\rho\sigma} \right)D_{\rho}\,a^{(1)}_{\sigma}
       \\
       &- \sqrt{2}\,a^{(0)}_{\sigma}  e^{-\phi_0}\left(2\bar{F}^{(0)\rho (\mu  }\bar{g}^{\nu) \sigma }-2\bar{F}^{(0)\sigma (\mu  }\bar{g}^{\nu) \rho } -\bar{g}^{\mu\nu}\bar{F}^{(0)\rho \sigma}\right)D_{\rho}h_{\mu\nu}\,
       \\
       & -\sqrt{2}\, a^{(1)}_\sigma e^{\phi_0}\left(2\bar{F}^{(1)\rho (\mu  }\bar{g}^{\nu) \sigma }-2\bar{F}^{(1)\sigma (\mu  }\bar{g}^{\nu) \rho } - \bar{g}^{\mu\nu }\bar{F}^{(1)\rho\sigma} \right)D_{\rho}h_{\mu\nu} 
       \\
       & -2\,a^{(0)}_\mu  e^{-\phi_0} \bar{F}^{(0)\rho \mu }D_{\rho}\phi +2\, a^{(1)}_\mu  e^{\phi_0} \bar{F}^{(1)\rho \mu}D_{\rho}\phi
       \\
      & + 2\, \phi\, e^{-\phi_0} \bar{F}^{(0)\rho\mu }D_{\rho}\,a^{(0)}_{\mu} -2\, \phi\, e^{\phi_0} \bar{F}^{(1)\rho\mu }D_{\rho}\,a^{(1)}_{\mu}\,.
       \end{split}
   \end{align}

\subsection*{Evaluation of Trace and Seeley-DeWitt coefficient}

Having the quadratic operators identified above, we compute the Seeley DeWitt coefficient $a_{4}(x)$ defined in \eqref{eq:seeley_short}. For this, we first evaluate ${\rm Tr}(I) $, ${\rm Tr}(E)$, ${\rm Tr}(E)^2$, and ${\rm Tr}(\Omega_{\rho\sigma}\Omega^{\rho\sigma})$,  where $I$, $E$, and $\Omega $ are respectively defined in  \eqref{eq:Imn}, \eqref{eq:connection_short} and  \eqref{eq:Omega_short}. For trace computation, we use the effective metric $I$ as commented in \eqref{rule_product} and \eqref{rule_trace}.
The trace of $I$ given in \eqref{eq:Imn} , is obtained as
\begin{eqnarray}\label{eq:trI}
     {\rm Tr }(I) = 10+4+4+1=19\,,
\end{eqnarray}
which corresponds to a total of 19 off-shell degrees of freedom: 10 from the graviton, 4 from each of the Maxwell fields, and 1 from the scalar field.

The traces of $E$, $E^2$ and $\Omega^2$ 
can be subsequently calculated as follows:
\begin{eqnarray}\label{eq:traceE}
    \text{Tr} (E)&=& 8\left( e^{-2\phi_0} (\bar F^{(0)})^2+ \; e^{2\phi_0} (\bar F^{(1)})^2\right)-3\;\bar R\,,
\end{eqnarray}
\begin{equation}\label{eq:traceE2}
\begin{aligned}
\text{Tr} (E^2)
&= 3\,\bar R_{\mu\nu\rho\sigma}\bar R^{\mu\nu\rho\sigma}
+ 3\Bigl(
e^{-2\phi_0}\,\bar F^{(0)\mu\nu}\bar F^{(0)\lambda\sigma}
+ e^{2\phi_0}\,\bar F^{(1)\mu\nu}\bar F^{(1)\lambda\sigma}
\Bigr)\bar R_{\mu\nu\lambda\sigma}\\[2pt]
&\quad+ \frac{7}{2}\,\bar R_{\mu\nu}\bar R^{\mu\nu}
- \frac{3}{8}\,\bar R^2
- 5\,\bar R\Bigl(
e^{-2\phi_0}(\bar F^{(0)})^2
+ e^{2\phi_0}(\bar F^{(1)})^2
\Bigr)\, ,
\end{aligned}
\end{equation}

\begin{equation}\label{eq:traceOmega2}
\begin{aligned}
\text{Tr}(\Omega_{\rho\sigma}\Omega^{\rho\sigma})
&= -8\,\bar R_{\mu\nu\rho\sigma}\bar R^{\mu\nu\rho\sigma}
- 18 \Bigl(
e^{-2\phi_0}\,\bar F^{(0)\mu\nu}\bar F^{(0)\lambda\sigma}
+ e^{2\phi_0}\,\bar F^{(1)\mu\nu}\bar F^{(1)\lambda\sigma}
\Bigr)\bar R_{\mu\nu\lambda\sigma}
\\[2pt]
&\quad
- 5\,\bar R_{\mu\nu}\bar R^{\mu\nu}
+ \frac{5}{4}\,\bar R^2
+ 16\,\bar R\Bigl(
e^{-2\phi_0}(\bar F^{(0)})^2
+ e^{2\phi_0}(\bar F^{(1)})^2
\Bigr)
\\[2pt]
&\quad
+ 32\,(\bar F^{(0)})^2(\bar F^{(1)})^2 \, .
\end{aligned}
\end{equation}
The detailed calculations of $E$, $\Omega$, and their traces are given in \ref{app:detailedcal}. To simplify the expressions, we utilize the identities for background fields given in Appendix~\ref{app:backidentites}.
Substituting the traces computed above into \eqref{eq:seeley_short}, we obtain the final Seeley-DeWitt coefficient $a_4$ as
\begin{equation}\label{eq:a4withoutghost}
\begin{aligned}
(4\pi)^2\, a^{\text{gauge-fixed}}_4
&= \frac{169}{180}\,\bar R_{\mu\nu\rho\sigma}\bar R^{\mu\nu\rho\sigma}
+ \frac{221}{180}\,\bar R_{\mu\nu}\bar R^{\mu\nu}
- \frac{23}{72}\,\bar R^2
\\[2pt]
&\quad
+ \frac{1}{6}\,\bar R\Bigl(
e^{-2\phi_0}(\bar F^{(0)})^2
+ e^{2\phi_0}(\bar F^{(1)})^2
\Bigr)
+ \frac{8}{3}\,(\bar F^{(0)})^2(\bar F^{(1)})^2 \, .
\end{aligned}
\end{equation}

\subsubsection{From ghost action}\label{sec:ghosts}

In addition to the gauge-fixed action \eqref{eq:Action_Phi} and \eqref{eq:gaugefix1}, we need to add the action of the ghost fields corresponding to the gauge-fixing of diffeomorphism and the two $U(1)$ symmetries \eqref{eq:gaugefix1}. Let us compute the corresponding Seeley DeWitt coefficient for this ghost action. 

 Denoting the antighost-ghost pairs as  $(b_\mu\,, c_\mu)$  corresponding to the diffeomorphism and $(b^{(I)}\,, c^{(I)})$  with $I=0,1$, corresponding to the two $U(1)$ gauge symmetries, the resulting ghost action take the form~\cite{Banerjee:2010qc}
\begin{eqnarray}
    S_{\text{ghost}}&=& \int d^4x \sqrt{ {g}}\left[ b^\mu \delta_{\text{brst}}\left(D^\rho h_{\mu\rho} - \tfrac{1}{2} D_\mu h^\rho_{\ \rho}\right) +b^{(I)} \delta_{\text{brst}} \left( D^\rho a^{(I)}_\rho \right) \right]\,.
 \end{eqnarray}
 Since the BRST transformations around the background geometry have
 \begin{equation}
     \delta_{\text{brst}}h_{\mu\nu}= D_\mu c_\nu + D_\nu c_\mu +\cdots\,, \qquad \delta_{\text{brst}}a^{(I)}_\mu = D_{\mu}c^{(I)} - 2 F^{(I)}_{\mu\nu}c^\nu + 2 D_\mu(A^{(I)}_\rho c^\rho)+\cdots\,,
 \end{equation}
 the quadratic terms in the ghost action are given by 
 \begin{equation}\label{eq:ghostS}
     S^{(2)}_{\text{ghost}}=\frac{1}{2}\int d^4x \sqrt{\bar{g}}\left[2 b_{\mu} \big( {\bar g}^{\mu\nu} D_\rho D^\rho + \bar R^{\mu\nu} \big) c_{\nu} 
  + 2 b^{(I)} D_\rho D^\rho c^{(I)}  
{\color{black}-} 4 {\bar F}^{(I)\rho\nu} b^{(I)} D_\rho c_{\nu}    \right]\,.
 \end{equation}
Again, since we will consider only quadratic action, all the fields except the ghost fields are in their background values, and thus we have used the on-shell equation $D^\mu \bar F_{\mu\nu}=0$.  We have further taken field redefinition $c^{(I)}\rightarrow c^{(I)}-2 \bar A^{(I)}_\rho c^\rho $ to arrive at the final form of the ghost action \eqref{eq:ghostS}. 

For the purpose of computing the heat kernel coefficient, we can treat $b_\mu\,, c_\mu$ as vector fields and $b^{(I)}\,, c^{(I)}$ as scalar fields instead of Grassmann variables. In order to make the kinetic term in \eqref{eq:ghostS} diagonal, we make the change of variables \cite{Charles:2015eha}
\begin{eqnarray}\label{eq:ghostredef}
    b_{\mu} &\;\;\longrightarrow\;\; \tfrac{1}{\sqrt{2}} \left( c^\prime _{\mu} - i\, b^\prime_{\mu} \right), 
    \qquad 
    c_{\mu} &\;\;\longrightarrow\;\; \tfrac{1}{\sqrt{2}} \left( c^\prime_{\mu} + i\, b^\prime_{\mu} \right), \nonumber\\
    b^{(I)} &\;\;\longrightarrow\;\; \tfrac{1}{\sqrt{2}} \left( c^{\prime (I)} - i\, b^{\prime (I)} \right), 
    \qquad 
    c^{(I)} &\;\;\longrightarrow\;\; \tfrac{1}{\sqrt{2}} \left( c^{\prime(I)} + i\, b^{\prime(I)} \right)\,.
\end{eqnarray}
We also adjust the total derivative terms to make the action hermitian, so that we find
\begin{eqnarray}
    \mathcal{S}^{(2)}_{\text{ghost}}&=&\frac{1}{2}\int d^4 x\sqrt{ \bar g}\;\bigg[b^\prime_\mu (\bar R^{\mu\nu} +\bar g^{\mu\nu}D^\rho D_\rho)b^\prime_\nu +c^\prime_\mu (\bar R^{\mu\nu}+\bar g^{\mu\nu}D^\rho D_\rho) c^\prime_\nu +b^{\prime (I)} D_\rho D^\rho b^{\prime (I)}\nonumber
     \\
     &&\qquad\qquad\qquad\quad +c^{\prime (I)} D_\rho D^\rho c^{\prime (I)}-  (b^{\prime (I)}+i c^{\prime (I)} ) \bar F^{(I)\rho\mu}D_\rho\left( \;b^\prime_\mu- i c^\prime_\mu\right)\nonumber
     \\
     &&\qquad\qquad\qquad\quad + (b^\prime_\mu- i c^\prime_\mu) \bar F^{(I)\rho\mu}D_\rho\left(  b^{\prime (I)}+i  c^{\prime(I)}\right)\bigg]\,.
\end{eqnarray}
From this quadratic action, we can identify the necessary matrices as defined in \eqref{eq:kineticop_short}, i.e, $I^{mn}\,, N_{\rho}^{mn}= 2(\omega_\rho)^{mn}\,,P^{mn}$ :
\begin{eqnarray}\label{eq:Ighost}
\xi_m I_{\text{ghost}}^{mn}D_\rho D^\rho\xi_n&=& b^\prime_\mu \bar g^{\mu\nu}\bar D_\rho D^\rho  b^\prime_\nu+ c^\prime_\mu \bar g^{\mu\nu} D_\rho D^\rho c^\prime_\nu+ b^{\prime (I)} D_\rho D^\rho b^{\prime (I)}\nonumber\\&&+c^{\prime (I)} D_\rho D^\rho c^{\prime(I)}\,,
\end{eqnarray}
\begin{eqnarray}
\label{eq:Pghost}
\xi_m P_{\text{ghost}}^{mn}\xi_n&=& b^\prime_\mu \bar R^{\mu\nu} b^\prime_\nu+ c^\prime_\mu \bar R^{\mu\nu} c^\prime_\nu\,,
\end{eqnarray}
\begin{align}\label{eq:omegaghost}
\begin{split}
     2\xi_m (\omega^\rho_{\text{ghost}})^{mn}D_\rho\xi_n 
    = &\; \;\; b'_{\mu}\bar F^{(I)\rho\mu}D_\rho b'^{(I)}- b'^{(I)}\bar F^{(I)\rho\mu}D_\rho b'_{\mu} 
    \\
    &+c'_{\mu}\bar F^{(I)\rho\mu}D_\rho c'^{(I)} - c'^{(I)}\bar F^{(I)\rho\mu}D_\rho c'_{\mu}
    \\
    & +ib'_{\mu} \bar F^{(I)\rho \mu} D_\rho c'^{(I)}-i c'^{(I)} \bar F^{(I)\rho \mu} D_\rho  b'_{\mu} 
    \\
    &  - ic'_{\mu} \bar F^{(I)\rho \mu} D_\rho b'^{(I)}  
    +  i b'^{(I)} \bar F^{(I)\rho \mu} D_\rho  c'_{\mu}\,.
\end{split}
\end{align}

Having the quadratic operator identified above, we can evaluate the $E^{mn}_{\rm ghost}$ and $(\Omega_{\rho\sigma})_{\text{ghost}}^{mn}$ defined in \eqref{eq:connection_short}  and \eqref{eq:Omega_short} as: 
\begin{eqnarray}
 \xi_m E^{mn}_{\text{ghost}}\xi_n= b^\prime_\mu \bar R^{\mu\nu} b^\prime_\nu+ c^\prime_\mu \bar R^{\mu\nu} c^\prime_\nu\;,
\end{eqnarray}
\begin{eqnarray}
    \xi_m (\Omega_{\rho\sigma})_{\text{ghost}}^{mn}\xi_n&=& b^\prime_\mu \bar R^{\mu\nu}_{\;\;~~{\rho\sigma}} b^\prime_\nu+c^\prime_\mu \bar R^{\mu\nu}_{\;\;~~{\rho\sigma}} c^\prime_\nu -\frac{1}{2}( b^\prime_{\mu}-ic^\prime_\mu)D^\mu \bar F^{(I)}_{\rho\sigma}(b^{\prime(I)}+i c^{\prime (I)})\nonumber\\&&+\frac{1}{2}(b^{\prime(I)} +i c^{\prime(I)} )D^\nu \bar F^{(I)}_{\rho\sigma} ( b^\prime_\nu-ic^\prime{_\mu})\,,
\end{eqnarray}
where we have used the Bianchi identity \eqref{eq:bianchi}, to simplify the expressions.

Now, to obtain the Seeley DeWitt coefficient $a_4$, we evaluate the trace ${\rm Tr}(I)$, ${\rm Tr}(E)$, ${\rm Tr}(E^2)$, and ${\rm Tr}(\Omega_{\rho\sigma}\Omega^{\rho\sigma}) $ to obtain the Seeley DeWitt coefficient:
The effective metric $I_{mn}^{\text{ghost}}$ on the ghost field space $(b^\prime_\mu,c^\prime_\mu,b^\prime_{(0)},b^\prime_{(1)},c^\prime_{(0)},c^\prime_{(1)})$ is
\begin{eqnarray}
    I^{\text{ghost}}_{mn}=\text{diag}\left(\bar g_{\mu\nu},\bar g_{\mu\nu},1,1,1,1\right )\,, 
\end{eqnarray}
and thus the trace is given by
\begin{eqnarray}
    \text{Tr}(I)=4+4+1+1+1+1=12\;.
\end{eqnarray}
The traces of $E$, $E^2$, and $\Omega^{\rho\sigma}\Omega_{\rho\sigma}$ are given by
\begin{equation}
    \text{Tr}(E)=2\bar R\,,\qquad  \text{Tr}(E^2)=2\bar R_{\mu\nu}\bar R^{\mu\nu}\,,\qquad \text{Tr}(\Omega_{\rho\sigma}\Omega^{\rho\sigma})=-2\bar R_{\mu\nu\lambda\kappa}\bar R^{\mu\nu\lambda\kappa}\,.
\end{equation}
Therefore, from \eqref{eq:seeley_short}, the Seeley-De Witt coefficient $a_4$ corresponding to the ghost contribution is given by
\begin{equation}\label{eq:a4_ghost}
    (4\pi)^2 a^{\text{ghost}}_4= \frac{1}{10}\bar R_{\mu\nu\lambda\sigma}\bar R^{\mu\nu\lambda\sigma} -\frac{1}{2}\bar R^2 -\frac{14}{15} \bar R_{\mu\nu}\bar R^{\mu\nu}\,.
\end{equation}
We have included an overall minus sign in the $a_{4}$ contribution of the ghost fields, reflecting the fact that they obey opposite spin statistics compared to the physical fields.

\subsection{Local contribution}
The total Seeley DeWitt coefficient  $a_4(x)$ combining the gauge fixed contribution \eqref{eq:a4withoutghost} to the ghost contribution  \eqref{eq:a4_ghost} evaluates to the following expression  
\begin{align}\label{eq:a4withghost}
(4\pi)^2\, a_4
&= \frac{187}{180}\,\bar R_{\mu\nu\rho\sigma}\bar R^{\mu\nu\rho\sigma}
 + \frac{53}{180}\,\bar R_{\mu\nu}\bar R^{\mu\nu}
 - \frac{59}{72}\,\bar R^2
\nonumber\\
&\quad
 + \frac{1}{6}\,\bar R\Bigl(
 e^{-2\phi_0}(\bar F^{(0)})^2
 + e^{2\phi_0}(\bar F^{(1)})^2
 \Bigr)
 + \frac{8}{3}\,(\bar F^{(0)})^2(\bar F^{(1)})^2 \, .
\end{align}
 Accordingly, the local contribution to the entropy function can be written as an integration over near horizon background,
\begin{eqnarray}
   C_{\text{local}}&=& \int^{{r}_c}_1 d {r} \int^{2\pi}_0 d \theta \int^{\psi_{max}}_0 d \psi \int^{2 \pi}_0 d\vartheta \sqrt{ \bar g} \,\,\, a_4\;\,,
\end{eqnarray}
 where $a_4$ is also evaluated on the near-horizon background.
 Since all the background invariants entering $a_4$ are independent of the coordinates $( r,\theta, \psi,\vartheta)$ (see \eqref{eq:backgroundinvr}), the $C_{\text{local}}$ reduces to the volume multiplied by $a_4$: 
\begin{eqnarray}
   C_{\text{local}}&=& a_4\int d^4 x\sqrt{ \bar g} \,\,\,\nonumber\\&=&  -2\pi \;L^2_1\;  L^2_2\; \mathbb{V}_{\mathbb{H}^2}\; a_4\,\,.
\end{eqnarray}
Here, following the spirit of presenting the entropy density as \eqref{eq:entopy}, we have used the regularized volume of AdS$_2$ as $-2\pi L_1^2$ and formally keep the volume of the hyperbolic surface as $L_2^2$ times the volume of the surface with unit radius $\mathbb{V}_{\mathbb{H}^2}$. 
Therefore, the final expression for the local contribution in terms of entropy density is given by 
\begin{eqnarray}\label{result:Clocal}
     \frac{C_{\text{local}}}{V_{\mathbb{H}^2}}= \frac{1}{\pi} \left(\frac{55}{144}-\frac{17}{60}\sinh^2{\nu}+\frac{37}{240}\tanh^2{\nu}\right)\,\,.
\end{eqnarray}
In accordance with the observation for the case of AdS black holes in \cite{David:2021eoq,Karan:2024gwf}, our result also includes the non-topological terms, as the parameter $\nu$ encodes the geometric data through the relation \eqref{eq:radius}. The novel point of our result is that the parameter $\nu$ is the unfixed horizon  modulus of the scalar field at the black hole horizon. Consequently, the logarithmic correction cannot be interpreted merely as a correction to the entropy, but instead plays the role of an effective quantum potential governing the integration over moduli in the quantum entropy function. We note that due to the negative sign in the $\sinh^2\nu$ term in \eqref{result:Clocal}, the integration over the horizon modulus $\nu$ is well-defined, and thus the modulus is stabilized due to the quantum lifting. \footnote{If we use the regularization of the hyperbolic horizon volume $V_{\mathbb{H}^2}$ in the same way as the volume of near horizon $V_{\text{AdS}_2} = -2\pi$, then the effective potential would have the wrong sign. However, this is not correct regularization, as otherwise the entropy obtained in \eqref{eq:entopy} would be negative definite.}  

We shall try the integration with the assumption of a trivial measure of $\nu$, 
\begin{equation}
    I(\alpha,\beta)\equiv \int_{-\infty}^{\infty}d\nu \exp\left( -\alpha \sinh^2\nu +\beta \tanh^2 \nu\right)\,.
\end{equation}
With the assumption $\alpha >0$, the integration is well-defined. Redefining the integration variable as $\sqrt{s}= \sinh\nu$, and taking the Taylor expansion in terms of $\beta$, we have
\begin{align}
    \begin{split}
         I(\alpha,\beta) &= e^{\beta} \sum_{n=0}^{\infty}\frac{(-\beta)^n}{n!}\int_0^\infty ds \, e^{-\alpha s} s^{-1/2}(1+s)^{-(n+1/2)}
         \\
         &= \sqrt{\pi }\sum_{n=0}^{\infty}\frac{(-\beta)^n}{n!}\,U(1/2, 1-n, \alpha)\,,
    \end{split}
\end{align}
where $U(x,y,z)$ is Tricomi's confluent hypergeometric function. For the case when $\beta=0$, the result is expressed in terms of the modified Bessel function
\begin{equation}
    I(\alpha,0)= e^{\alpha/2}K_0 (\frac{\alpha}{2})\,.
\end{equation}
In large $\alpha$ this result behaves as $I(\alpha,0)\sim \alpha^{-1/2}$.
\subsection{Global contribution}\label{sec:ZM}
Let us find the global contribution in \eqref{eq:Clocal}. As explained in \eqref{eq:beta}, the contribution to $C_{\text{global}}$ is only from graviton zero modes. As the building blocks of graviton zero modes, we shall first consider the one-form zero modes on the Euclidean AdS$_2$ as well as the hyperbolic space $\mathbb{H}^2$, and then combine them to find the number of graviton zero modes on AdS$_2 \times \mathbb{H}^2$.  A rank two tensor also has zero modes on those spaces; however, we will shortly see that it is not a relevant building block for the four-dimensional graviton zero modes.

\subsubsection*{Counting zero modes in AdS$_2$ and $\mathbb{H}^2$}
The Euclidean AdS$_2$ and $\mathbb{H}^2$  geometries are locally equivalent, and the metric is given by
\begin{equation}
    ds^2 = L^2 (d\eta^2 + \sinh^2\eta \,d\theta^2)\,.
\end{equation}
(From the AdS$_2$ metric given in \eqref{eq:NHmetric}, we can redefine $r= \cosh\eta$.)
However, we shall treat the zero modes on those backgrounds differently. 

On AdS$_2$, the one-form zero modes in a normalizable basis are given in the form of pure gauge modes as
\begin{equation}
    a_{\text{zm}}^{(\ell)} = d \varphi^{(\ell)}\,,\qquad \varphi^{(\ell)}= \frac{1}{\sqrt{2\pi |\ell|}} \left[\frac{\sinh\eta}{1+\cosh\eta} \right]^{|\ell|}e^{i\ell \theta}\,, \quad\ell = \pm 1\,, \pm 2\,, \pm3\,, \cdots, 
\end{equation}
where the scalar functions $\varphi^{(\ell)}$ are non-normalizable, so that those modes are not gauge redundancy. The modes satisfy the orthonormality
\begin{equation}
    \int d^2x \sqrt{g}\,\partial_m \varphi^{(\ell)\ast}\partial^m \varphi^{(\ell)} = \delta_{\ell \ell'}\,.
\end{equation}
The basis also satisfies
\begin{equation}
    \sum_\ell \partial_m \varphi^{(\ell)\ast}(x)\partial^m \varphi^{(\ell)}(x)  = \frac{1}{2\pi L^2}\,.
\end{equation}
This can be derived using the fact that due to the homogeneity of AdS$_2$, the sum is independent of the coordinate $x$ and hence can be evaluated at $\eta=0$, and then at $\eta=0$ only the $\ell = \pm 1$ terms contribute to the sum. The total number of zero modes is given by
\begin{equation}
    n^{\text{zm}}_{1,\text{AdS}_2} = \int_{\text{AdS}_2} d^2 x \sqrt{g} \sum_\ell \partial_m \varphi^{(\ell)\ast}(x)\partial^m \varphi^{(\ell)}(x) = \frac{1}{2\pi }\int^{\eta_0}_{1}d\eta \sinh\eta\int_0^{2\pi} d\theta = \cosh \eta_0 -1\,,
\end{equation}
where the final volume integral has the divergence, and we regularize it using holographic renormalization. Then we obtain
\begin{equation}
     n^{\text{zm}}_{1,\text{AdS}_2} = -1\,.
\end{equation}
On $\mathbb{H}^2$, the one-form has the same zero modes as the AdS$_2$. However, let us not use the holographic renormalization scheme to regularize the volume, and follow the spirit of presenting entropy density as in \eqref{eq:entopy}. The number of the zero modes of the one-form on $\mathbb{H}^2$ is thus given by
\begin{equation}
    n^{\text{zm}}_{1, \mathbb{H}^2} = \frac{1}{2\pi}\mathbb{V}_{\mathbb{H}^2}\,,
\end{equation}
i.e., the number density of the zero modes is given by $1/2\pi$. 
\subsubsection*{Zero modes on AdS$_2 \times \mathbb{H}^2$}
Let us split the  four-dimensional coordinate $x^\mu = \{x^m \,, y^\alpha\}$ for AdS$_2$ and $\mathbb{H}^2$ respectively, which splits the Laplace operator into 
\begin{equation}\label{eq:spltiOperator}
    D_\rho D^\rho = D_m D^m + D_\alpha D^\alpha\,. 
\end{equation}
We also consider the splitting the graviton modes $h_{\mu\nu}$ into $\{h_{mn}\,, h_{m\alpha}\,,h_{\alpha\beta} \}$. 

If the operator \eqref{eq:spltiOperator} acts on $h_{mn}(x,y)$, the $\mathbb{H}^2$ part acts as the scalar Laplacian $D_\alpha D^\alpha$, and thus the zero eigenvalue arises from the rank two tensor zero modes on AdS$_2$ 
\begin{equation}
 \sum_\ell   h_{mn}^{\text{zm}(\ell)}(x) \, c_{\ell}\,,
\end{equation}
with constants $c_{\ell}$. However, since a constant along $\mathbb{H}^2$ direction is not normalizable, there is no zero mode in the $h_{mn}$ component of the graviton.  Likewise, since a constant is not normalizable on AdS$_2$, there is no zero mode in $h_{\alpha\beta}$ component of the graviton. 

Now, we consider the mixed component of the graviton $h_{m\alpha}$. On the AdS$_2$ perspective, those modes form  non-abelian vector fields
\begin{equation}
    h_{m\alpha}(x,y) = a_{m}^{\;\;i}(x)k_{i \alpha}(y)
\end{equation}
where $k_{i\alpha}(y)\,, i =0,\pm 1$ is the Killing vector along $\mathbb{H}^2$ satisfying the Killing vector equation 
$\nabla_\alpha k_{i\beta}+ \nabla_\beta k_{i\alpha}=0$ that are $SL(2,R)$ generators. However,  the Killing vectors $k_{i\alpha}$ are not normalizable on the hyperbolic space. The same is true for the killing vectors on AdS$_2$. The only possible zero modes consist of the product of one-form zero modes on AdS$_2$ and $\mathbb{H}^2$ as
\begin{equation}
    h^{\text{zm}}_{m\alpha}(x,y) =\sum_{\ell, \ell'} c_{\ell \ell'}\, \partial_m \varphi^{(\ell)}(x)\,\partial_\alpha\varphi^{(\ell')}(y)\,.
\end{equation}
These are normalizable both on the AdS$_2$ and $\mathbb{H}^2$. Therefore, the total number of zero modes in the graviton is the product of the number of 1-form zero modes on AdS$_2$ and the number of 1-form zero modes on $\mathbb{H}^2$ as,
\begin{equation}
    n^{\text{zm}}_g = n^{\text{zm}}_{1, \text{AdS}_2} \times n^{\text{zm}}_{1, \mathbb{H}^2} = - \frac{1}{2\pi }\mathbb{V}_{\mathbb{H}^2}\,.
\end{equation}

Therefore, since the $\beta_g =2$,  the global contribution $C_{\text{global}}$ in \eqref{eq:Clocal}  to the 1-loop correction \eqref{eq:DeltaS} is given by
\begin{equation}
    C_{\text{global}} =  - \frac{1}{2\pi }\mathbb{V}_{\mathbb{H}^2}\,.
\end{equation}
\section{Reduction to simpler models}\label{sec:Reduction}
In this section, we present consistency checks of the one-loop result obtained in the previous section by considering several reductions of the theory at the level of quadratic fluctuation operators.   

Our one-loop result explicitly depends on the horizon moduli $\nu$, which parametrizes different classical backgrounds within the theory. However, since $\nu$ is a background parameter rather than a parameter of the theory, setting $\nu$ to a particular value does not interpolate between different theories. In particular, although setting $\nu=0$ yields a black hole solution that is also admitted by simpler theories such as minimal gauged supergravity or Einstein–Dilaton–Maxwell theory without a scalar potential, the nontrivial scalar–gauge interactions and potential terms that distinguish our theory from these simpler models remain present at the level of fluctuations. 
As a result, a naive $\nu \rightarrow 0$ limit of the final one-loop expression does not reproduce the corresponding results in those theories. 

For this reason, our strategy is to implement truncations directly at the level of the quadratic fluctuation operator. This allows us to obtain the appropriate quadratic operators of the simpler theories and to verify that the general structure of the Seeley–DeWitt coefficients is correctly reproduced. As a byproduct of this analysis, we also obtain the corresponding logarithmic corrections to the entropy of hyperbolic black holes in these truncated theories, which, to our knowledge, have not been previously studied.

In the following, we consider truncations to Einstein–Dilaton–Maxwell theory with a cosmological constant, minimal gauged supergravity, and pure gravity with a cosmological constant.

\subsection{Einstein–Dilaton–Maxwell theory with cosmological constant} 
The closest theory to which our gauged supergravity model can be reduced is the EDM theory with a cosmological constant, whose action is given by
\begin{eqnarray}
 S=\int d^4 x \sqrt{g} \left(R-e^{-2\Phi}F^{(0)}_{\mu\nu}F^{(0)\mu\nu}-e^{2\Phi}F^{(1)}_{\mu\nu}F^{(1)\mu\nu}-2D_\mu\Phi D^\mu\Phi- 2\Lambda\right)\,.
\end{eqnarray}
Note that the only difference from our action \eqref{eq:Action_Phi} is that the scalar potential $V(\Phi)$ is replaced by the cosmological constant $\Lambda$. This replacement results in the difference to the scalar equation of motion \eqref{eq:scalareom} by $V(\Phi)' =0$. This equation of motion of scalar together with those of graviton and gauge fields admit the black hole horizon solution \eqref{eq:NHmetric}, \eqref{eq:radius}, \eqref{eq:Fluxsolution}, and \eqref{eq:tausolution} with zero moduli parameter $\nu=0$. 

Consequently, around this near-horizon background,  the effective mass term of the scalar does not vanish as opposed to the case of our gauged supergravity because there is no $V''$ term in \eqref{eq:no_Mass}. Therefore, there is an additional mass term of the scalar
\begin{equation}
    \phi\left( e^{-2\phi_0} (\bar F^{(0)})^2+ e^{2\phi_0} (\bar F^{(1)})^2\right)\phi
\end{equation}
 to the $P^{mn}$ in \eqref{eq:Pmn}, which contributes to ${\rm Tr} (E)$ in \eqref{eq:traceE} and ${\rm Tr}(E^2)$ in \eqref{eq:traceE2}. The resulting Seeley DeWitt coefficient from the gauge fixed action takes the form 
\begin{align}
    (4 \pi)^2 a_4&= \frac{169}{180} \bar R_{\mu\nu\rho\sigma} \bar R^{\mu\nu\rho\sigma} +\frac{131}{180} \bar R_{\mu\nu} \bar R^{\mu\nu}-\frac{7}{36}\bar R^2+\frac{8}{3}(\bar F^{(0)})^2(\bar F^{(1)})^2\;.
\end{align}
 If we take $\phi_0 \rightarrow 0$ or absorb the $e^{\pm \phi_0}$ into the definition of $F^{(I)}_{\mu\nu}$ so that  $(\bar{F}^{(0)})^2 = (\bar{F}^{(1)})^2$, then using the identity
\begin{eqnarray}\label{eq:identity2}
 (\bar F^{(0)})^2(\bar F^{(1)})^2 =\frac{1}{4}\bar R_{\mu\nu}\bar R^{\mu\nu}-\frac{1}{16}\bar R^2
\end{eqnarray}
taken from \eqref{ruleF2=R2}, 
we can express the $a_4$ solely in terms of curvature invariant as

\begin{equation}
    (4\pi)^2 a^{\rm gauge   fixed}_4= \frac{169}{180} \bar R_{\mu\nu\rho\sigma} \bar R^{\mu\nu\rho\sigma}+\frac{251}{180}\bar R_{\mu\nu} \bar R^{\mu\nu}-\frac{13}{36}\bar R^2.
\end{equation}
\\
This result agrees with \cite{Karan:2022dfy}. To this gauged fixed contribution, we add the ghost contribution, which remains unchanged and is given by \eqref{eq:a4_ghost},  since the gauge-fixing conditions used in the present truncation remain the same.\footnote{ Our work corrects the ghost contributions of \cite{Karan:2022dfy}. } Therefore, the total Seeley- DeWitt coefficient for EDM theory is 
\begin{eqnarray}
    (4\pi)^2 a^{\rm EDM}_4&=& \frac{187}{180} \bar R_{\mu\nu\rho\sigma} \bar R^{\mu\nu\rho\sigma}+\frac{83}{180}\bar R_{\mu\nu} \bar R^{\mu\nu}-\frac{31}{36}\bar R^2\,.
\end{eqnarray}

On the near-horizon background of the hyperbolic BPS black hole, this Seeley DeWitt coefficient results in the local contribution as 
\begin{equation}
    \frac{C_{\text{local}}^{\text{EDM}}}{\mathbb{V}_{\mathbb{H}^2}} = \frac{101}{288 \pi}\;.
\end{equation}

\subsection{$\mathcal{N}=2$ minimal gauged supergravity (bosonic)}
Let us consider further reduction to the Einstein-Maxwell theory with cosmological constant, which is the bosonic truncation of minimal $\mathcal{N}=2$ supergravity. 

The action of this theory takes the following form,
\begin{eqnarray}
 S=\int d^4 x \sqrt{g} \left(R-F_{\mu\nu}F^{\mu\nu}- 2\Lambda\right)\,.
\end{eqnarray}
This is a consistent truncation of our gauged supergravity model as the black hole horizon solution \eqref{eq:NHmetric}, \eqref{eq:radius}, and \eqref{eq:Fluxsolution} with zero moduli parameter $\nu=0$ is also a solution of this theory. 

\par To obtain a consistent truncation to this theory, we
freeze the scalar to its background value such that the scalar degree of freedom vanishes and the potential becomes the cosmological constant, $V_0\rightarrow \Lambda$.

We also freeze one of the gauge field degrees of freedom by identifying 
\begin{equation}
    e^{-\phi_0}A^{(0)}_\mu = e^{\phi_0}A^{(1)}_\mu =\frac{1}{\sqrt{2}}\tilde{A}_\mu\,,
\end{equation}

together with switching off one of the ghost fields. Thus, reducing the number of propagating degrees of freedom from four to two. 
With these truncation, the operator $P^{mn}$ in \eqref{eq:Pmn} and $\omega^\rho$ in \eqref{eq:omega} reduce to 
 \begin{equation}
    \begin{split}
     \xi_m P^{mn}\xi_n&=h_{\mu\nu} \bigg\{I^{\mu\nu,\lambda\sigma}\left(\frac{1}{2}\bar R- ( \tilde F)^2\right)+4\tilde F^{\mu(\lambda}\tilde F^{\sigma)\nu}-2\bar R^{\mu(\lambda\sigma)\nu}\\&\qquad\qquad-\bar R^{\mu(\lambda}\bar g^{\sigma)\nu}-\bar R^{\nu(\lambda}\bar g^{\sigma)\mu}\bigg\}h_{\lambda\sigma}\\&\quad-\tilde a_\mu  \bar R^{\mu\nu}\tilde a_\nu -\sqrt{2} h_{\mu\nu} D^{(\mu}\tilde F^{\nu)\sigma}\tilde a_\sigma-\sqrt{2} \tilde a_\sigma D^{(\mu}\tilde F^{\nu)\sigma}h_{\mu\nu}\,,
    \end{split}
\end{equation}
\begin{equation}
    \begin{split}
       2 \xi_m (\omega^\rho)^{mn} \xi_n& =  \sqrt{2}\;h_{\mu\nu} \left(2 \tilde F^{\rho(\mu}\bar g^{\nu)\sigma}-2 \tilde F^{\sigma(\mu}\bar g^{\nu)\rho}-\bar g^{\mu\nu} \tilde F^{\rho\sigma}
\right)\tilde a_\sigma\\&-\sqrt{2}\;\tilde a_\sigma \left(2 \tilde F^{\rho(\mu}\bar g^{\nu)\sigma}-2 \tilde F^{\sigma(\mu}\bar g^{\nu)\rho}-\bar g^{\mu\nu} \tilde F^{\rho\sigma}
\right)h_{\mu\nu}\,.
    \end{split}
\end{equation}
This truncation is followed by the reduction of matrices $E^{mn}$ and $\Omega_{\rho\sigma}^{mn}$, and then we obtain the Seeley DeWitt coefficient by computing the relevant traces as follows.
The field space identity $I^{mn}$ 
receives contributions only from the graviton and a single graviphoton. As a result, its trace is reduced to $14$ as  
\begin{equation}
    \text{Tr}(I)= 10+4=14\,.
\end{equation}
The trace of $E$, $E^2$ and $\Omega_{\rho\sigma}\Omega^{\rho\sigma}$ are
\begin{equation}
    \text{Tr}(E)= 6(\tilde F)^2-2 \bar R\;,
\end{equation}
\begin{eqnarray}
    \text{Tr}(E^2)&=&3 \bar R_{\mu\nu\rho\sigma}\bar R^{\mu\nu\rho\sigma}+3 \tilde F^{~\mu\nu}\tilde F^{~\rho\sigma}\bar R_{\mu\nu\rho\sigma}-7 \bar R_{\mu\nu}\bar R^{\mu\nu}+ 2 \bar R^2\nonumber\\&&- \frac{9}{2} \bar R\,(\tilde F)^2+9 (\tilde F)^2\;,
\end{eqnarray}

\begin{eqnarray}
    \text{Tr}(\Omega_{\rho\sigma}\Omega^{\rho\sigma})&=&-7 \bar R_{\mu\nu\rho\sigma}\bar R^{\mu\nu\rho\sigma}-18 \tilde F^{~\mu\nu}\tilde F^{~\rho\sigma}\bar R_{\mu\nu\rho\sigma} +56 \bar R_{\mu\nu}\bar R^{\mu\nu}-14 \bar R^2\nonumber\\&&-54(\tilde F)^2+15\bar R (\tilde F)^2\;.
\end{eqnarray}
Therefore, substituting these trace data  in \eqref{eq:seeley_short}, we obtain the $a_4$ for the gauge fixed action,
\begin{equation}
    (4\pi)^2a_4^{\text{gaugefixed}} =  \frac{179}{180}\,\bar  R_{\mu\nu\rho\sigma}\bar R^{\mu\nu\rho\sigma}+ \frac{49}{45}\, \bar R_{\mu\nu}\bar R^{\mu\nu}-\frac{11}{36}\, \bar R^2  \,.
\end{equation}
The ghost contribution comes from the ghosts corresponding to the diffeomorphism and to the $U(1)$ gauge symmetry. 
This contribution is \cite{Charles:2015eha,David:2021eoq,Karan:2021teq}
\begin{equation}
     (4\pi)^2a^{\text{ghost}}_4 = \frac{1}{9}\bar R_{\mu\nu\rho\sigma}\bar R^{\mu\nu\rho\sigma}-\frac{17}{18} \bar R_{\mu\nu}\bar R^{\mu\nu} -\frac{17}{36} \bar R^2.
\end{equation}
 
Therefore, the net bosonic contribution for the complete bosonic sector of minimal $\mathcal{N}=2$ is given by,
\begin{equation}\label{eq:a4Minimal}
    (4\pi)^2a^{\rm B}_4 =\frac{199}{180} \bar R_{\mu\nu\rho\sigma}\bar R^{\mu\nu\rho\sigma}+ \frac{13}{90}\bar R_{\mu\nu} \bar R^{\mu\nu}- \frac{7}{9}\, R^2.
\end{equation}
This matches with the bosonic sector of minimal $\mathcal{N}=2$ gauged supergravity discussed in~\cite{David:2021eoq}. The corresponding local contribution to the entropy of the hyperbolic BPS black hole is

\begin{equation}
    \frac{C_{\text{local}}}{\mathbb{V}_{\mathbb{H}^2}}= \frac{5}{18\pi}\;.   
\end{equation}
\subsection{Gravity with a cosmological constant}
We can consider further reduction to pure gravity with cosmological constant, whose action is given by
\begin{eqnarray}
 S=\int d^4 x \sqrt{g} \left(R- 2\Lambda\right)\,.
\end{eqnarray}
This is a consistent truncation of our gauged supergravity model, as the black hole near-horizon background \eqref{eq:NHmetric} and \eqref{eq:radius} with $\nu=0$ is also a solution of this theory.

For this truncation, we keep only the graviton fluctuations, freezing out all the other degrees of freedom. As a consequence,
$\omega^\rho$ vanishes,  the matrix
$P^{mn}$ receives contribution only from \eqref{eq:EHvariation}, and thus becomes equal to $E^{mn}$,

\begin{equation}
    \xi_m P^{mn} \xi_n=\xi_m E^{mn} \xi_n=h_{\mu\nu}( -\Lambda g^{\mu\nu} g^{\lambda\sigma}-2 \bar R^{\lambda(\mu\nu)\sigma})h_{\lambda\sigma}\,,
\end{equation}
where we have used $\bar R_{\mu\nu}= g_{\mu\nu}\Lambda$ to simplify the expression. The effective metric $I^{mn} $ is the De-Witt metric $I^{\mu\nu,\lambda\sigma}$ coming from the graviton. Thus, its trace is
 \begin{eqnarray}
     \text{Tr} I= 10\,.
 \end{eqnarray}
The trace structures required for the computation of  $a_4$ can be expressed as,
\begin{align}
&\text{Tr}(E)= -4\Lambda\;,\\&
\text{Tr}(E^2)= 16\Lambda^2-4 \bar R_{\mu\nu} \bar R^{\mu\nu}+3 \bar R_{\mu\nu\rho\sigma}\bar R^{\mu\nu\rho\sigma}\,,\\&
\text{Tr}(\Omega_{\rho\sigma}\Omega^{\rho\sigma})=-6\bar R_{\mu\nu\rho\sigma}\bar R^{\mu\nu\rho\sigma}\;.
\end{align}
Combining these trace data, the $a_4$ for the gauge fixed action becomes, 
 \begin{equation}
   (4 \pi)^2  a_4^{\text{gauge fixed}}= - \frac{2}{3}\Lambda^2 +\frac{19}{18}\bar R_{\mu\nu\rho\sigma}\bar R^{\mu\nu\rho\sigma}\,.
 \end{equation}
 
 The ghost contribution arises corresponding to the diffeomorphism.
 The $E^{mn}$ contribution is 
 \begin{eqnarray}
     \xi_m E_{\rm ghost}^{mn}\xi_n= b_\mu \bar R^{\mu\nu} b_\nu+ c_\mu \bar R^{\mu\nu} c_\nu\,,
 \end{eqnarray}
 whereas the effective connection $\Omega$ is
 \begin{equation}
     \xi_m( \Omega_{\rho\sigma})_{\rm ghost}^{mn}\xi_n= b_\mu \bar R^{\mu\nu}_{~~~~~\;\rho\sigma} b_\nu+ c_\mu \bar R^{\mu\nu}_{~~~~~\;\rho\sigma} c_\nu\,.
 \end{equation}
The trace of $I$ operator for the two vector ghosts $b_\mu$ and $c_\mu$ is
\begin{equation}
    \text{Tr}(I)= 4+4=8\,.
\end{equation}
Therefore, the $a_4$ contribution from the ghost sector comes out to be 
 \begin{equation}
   (4 \pi)^2  a_4^{\text{ghost}}= -\frac{164}{15} \Lambda^2 +\frac{11}{90}\bar R_{\mu\nu\rho\sigma}\bar R^{\mu\nu\rho\sigma}\,.
 \end{equation}
 
The total Seeley-De Witt coefficient for the complete theory is given by
\begin{equation}
   180(4 \pi)^2  a_4^{\text{Gravity}}=- 2088 \Lambda^2 +212\bar R_{\mu\nu\rho\sigma}\bar R^{\mu\nu\rho\sigma}\,,
 \end{equation}
which agrees with \cite{CHRISTENSEN1980480}.
The corresponding local contribution to the entropy of the hyperbolic black hole is
\begin{eqnarray}
    \frac{C^{\rm Gravity}_{\rm local}}{\mathbb{V}_{\mathbb{H}^2}}=\frac{229}{1440\pi}\;.
\end{eqnarray}

\section{Discussion and outlook}\label{sec:conclusion}
In this work, we studied one-loop logarithmic corrections to the entropy of static hyperbolic BPS black holes in asymptotically AdS$_4$ spacetime within Fayet–Iliopoulos gauged supergravity. Our analysis was performed in a consistent real-scalar truncation of the $\mathcal{N}=2$ theory specified by the prepotential $F=-\mathrm{i}X^0X^1$, which corresponds to an Einstein–Dilaton–Maxwell theory with a nontrivial scalar potential. At the classical level, this system exhibits flat directions in the attractor equations, leading to continuous families of horizon moduli while the Bekenstein–Hawking entropy remains purely charge-dependent.

The logarithmic correction was computed using the heat kernel method, incorporating both local contributions determined by the Seeley–DeWitt coefficients and global contributions arising from zero modes. In particular, the analysis of zero modes for hyperbolic horizons requires a treatment distinct from the case of compact horizons, which we have carried out explicitly.

A central outcome of our analysis is that the one-loop logarithmic correction includes not only topological but also non-topological contributions, as anticipated in previous studies of AdS black holes \cite{David:2021eoq,Karan:2024gwf}. In our case, the non-topological terms arise through the dependence on the parameter $\nu$, which encodes geometric data of the near-horizon background. A distinctive feature of the present setup is that $\nu$ is not merely a background parameter, but precisely the unfixed scalar modulus at the black hole horizon. Consequently, the logarithmic correction cannot be regarded simply as a constant shift of the entropy; rather, it enters the quantum entropy function as an effective potential governing the integration over the horizon modulus. 

Importantly, the structure of the local contribution implies that the effective potential is bounded from below, rendering the moduli integral well-defined and dynamically lifting the classical flat direction. In this sense, the one-loop effect stabilizes the horizon modulus at the quantum level and, to our knowledge, provides the first explicit example in which quantum corrections are shown to dynamically lift classical attractor flat directions in AdS gauged supergravity.

To further support the robustness of our analysis, we performed several nontrivial consistency checks of the one-loop computation. By considering appropriate truncations of the theory and the quadratic fluctuation operator, we verified that the general structure of the Seeley DeWitt coefficients is correctly reproduced in simpler settings, including Einstein–Dilaton–Maxwell theories without scalar potentials, minimal gauged supergravity, and pure gravity. The corresponding logarithmic corrections to the hyperbolic BPS black holes in these truncated theories have not been previously explored.

Several important directions remain open for future investigation. A natural next step is to extend the present analysis to the fully supersymmetric theory and determine whether additional cancellations arise once the complete supermultiplet spectrum is taken into account. Such an extension would clarify the role of supersymmetry in the quantum lifting mechanism identified here and may provide further insight into its microscopic interpretation. In particular, a fully supersymmetric treatment could lead to nontrivial constraints or predictions that can be compared with microscopic descriptions of AdS black hole entropy~\cite{Cabo-Bizet:2017jsl}.

Although a universal analysis is generally not possible in gauged supergravity, it would also be interesting to explore other choices of prepotentials and black hole backgrounds to assess the generality of the mechanism uncovered in this work. Such investigations may help clarify the relation between logarithmic corrections in AdS black holes and recent developments in supersymmetric localization \cite{Hristov:2018lod,Hristov:2019xku}, and guide further applications of localization techniques in supergravity \cite{Dabholkar:2010uh,Dabholkar:2011ec,Jeon:2018kec,Iliesiu:2022kny,Dabholkar:2014wpa,Nian:2017hac,Hristov:2021zai}.

Finally, our results suggest a possible connection between quantum corrections and the integration measure over unfixed horizon moduli in the quantum entropy function. The one-loop effective potential generated by logarithmic corrections may provide a dynamical input relevant for the moduli integration. Determining the correct measure from first principles and understanding its relation to microscopic descriptions of AdS black holes remain open problem for future study~\cite{LopesCardoso:2022hvc}.
In our setup, the horizon modulus $\nu$ corresponds, a priori, to a non-normalizable scalar mode in AdS$_2$, whose quantum lifting renders the path integral well-defined. It would be interesting to explore whether this phenomenon is related to the role of non-normalizable scalar modes required by supersymmetry in AdS backgrounds~\cite{GonzalezLezcano:2023cuh,GonzalezLezcano:2023uar,GonzalezLezcano:2024rsi}.


\acknowledgments
 B.P. acknowledges CSIR-HRDG, Govt of India, for financial support received through Grant No. 03WS(003)/2023-24/EMR-II/ASPIRE. This work was also partially supported by Asia Pacific Center for Theoretical Physics (APCTP). We wish to specially thank Ashoke Sen for reading the manuscript and for his valuable comments. We also thank Abhinava Bhattacharjee, Alfredo González Lezcano,  Sudip Karan, Robert de Mello Koch, Finn Larsen, Sameer Murthy, Leopoldo A. Pando Zayas,   Augniva Ray, and Xuao Zhang and  for valuable discussions. 
 I.J. acknowledges the warm hospitality of Heribertus Bayu Hartanto, Hyun-Sik Jeong, Kanghoon Lee,  Matthew Roberts, and Junggi Yoon at APCTP. A.S. acknowledges Indian Institute of Technology Bombay for the Institute postdoctoral fellowship. 

\appendix

\section{Review 
}\label{app:N=2SUGRA}

\subsection{${\cal N}=2$ FI gauged supergravity}\label{sec:review}
We shall briefly review the $\mathcal{N}=2$, $D=4$ Fayet-Iliopoulos(FI) gauged supergravity coupled to Abelian vector multiplets \cite{Cacciatori:2004rt,Cacciatori:2008ek,Cacciatori:2009iz,VanProeyen:2004xt}. The bosonic sector of the theory consists of the vierbein $e^a_{\ \mu}$, a set of gauge fields $A^I_\mu$ with $I=0,\dots,n_V$, and complex scalar fields $z^\alpha$ ($\alpha=1,\dots,n_V$). The physical scalars are defined through $z_{}^\alpha =\frac{ X^\alpha}{X^0}$, where $X^I$ are homogeneous (projective) complex coordinates on the special Kähler manifold.
The scalar fields parametrize a special Kähler manifold, with its geometry encoded in the covariantly holomorphic symplectic section 
\begin{equation}\label{eq:holomorph}
    \mathcal{V} =
    \begin{pmatrix}
        X^I \\ F_I
    \end{pmatrix}\;, 
   { \qquad F_I \equiv \frac{\partial F(X)}{\partial X^I}}\;,
\end{equation}
where $F(X)$ is the holomorphic prepotential, a homogeneous function of degree two.
It is often convenient to parametrize the holomorphic sections as $X^I = y \, Z^I$, with $y = e^{K(z,\bar{z})/2}$. In this parametrization, all dependence on $X^I$ can be expressed in terms of $Z^I$, and the symplectic section takes the form \cite{Cacciatori:2009iz}
\begin{equation} \label{eq:holomorphicsec}
    \mathcal{V} =
    \begin{pmatrix}
        X^I \\ F_I
    \end{pmatrix} 
    = e^{K(z,\bar{z})/2} 
    \begin{pmatrix}
        Z^I(z)\\ \partial_{Z^I} F(Z)
    \end{pmatrix}
    = e^{K(z,\bar{z})/2} v(z)\;,
\end{equation}
where $v(z) = (Z^I(z), \, \partial_{Z^I} F(Z))^T$ is the symplectic vector. 
The Kähler potential $K(z,\bar z)$ is defined through
\begin{equation}
    e^{-K(z,\bar z)} = -i \langle v(z), \bar v(\bar z)\rangle 
    \equiv -i\, v(z)^T 
    \begin{pmatrix} 
        0 & 1 \\ -1 & 0 
    \end{pmatrix} \bar{v}(\bar{z}) = i (\bar{Z}^I F_I - Z^I \bar{F}_I)%
    \,.
\end{equation}
From this the Kähler metric for the scalar field $z^\alpha \,, \bar{z}^{\bar{\alpha}}$ is obtained as 
\begin{equation}\label{eq:kahlermetdef}
    g_{\alpha\bar{\beta}} = \partial_\alpha \partial_{\bar{\beta}} K \,.
\end{equation}

The gauge couplings of the theory are encoded in the period matrix $\mathcal{N}_{IJ}$, given by \cite{Freedman:2012zz}
\begin{equation}
  \mathcal{N}_{IJ} 
  = \bar{F}_{IJ} 
  + \frac{i\, N_{IK} X^K N_{JL} X^L}{N_{MN} X^M X^N}\;,
\end{equation}
where $N_{IJ} \equiv 2\, \mathrm{Im}\, F_{IJ} = -i F_{IJ} + i \bar{F}_{IJ}$, and $F_{IJ}$ denotes the second derivative of the prepotential $F$.  
A scalar potential is generated through $U(1)$ Fayet–Iliopoulos gauging \cite{Cacciatori:2009iz,VanProeyen:2004xt,Astesiano:2020fwe,Cacciatori:2008ek}:
\begin{equation}\label{eq:generalV}
    V = -2 g^2 \, \xi_I \xi_J \left[ (\mathrm{Im}\, \mathcal{N})^{-1|\,IJ} + 8 \bar X^I X^J \right],
\end{equation}
where $g$ is the gauge coupling and $\xi_I$ are the FI parameters and defining $g_I = g \xi_I$. 

The bosonic Lagrangian of the theory then reads
\begin{equation}\label{eq:generalL}
    \mathcal{L} = \frac{1}{4} (\mathrm{Im}\, \mathcal{N})_{IJ} F^I_{\mu\nu} F^{J \mu\nu} 
    - \frac{1}{8} (\mathrm{Re}\, \mathcal{N})_{IJ} \epsilon^{\mu\nu\rho\sigma} F^I_{\mu\nu} F^{J \mu\nu} 
    + \frac{R}{16 \pi G} 
    - g_{\alpha \bar{\beta}} \, \partial_\mu z^\alpha \partial^\mu \bar{z}^{\bar{\beta}} 
    - V(\tau)\;.
\end{equation}
\subsection{The $F= -iX^0 X^1$ model and its consistent truncation}
We focus on a model involving a single abelian vector multiplet coupled to the gravity multiplet, specified by the following holomorphic prepotential \cite{Cacciatori:2009iz,Chimento:2015rra}:
\begin{equation}\label{eq:prepotential}
    F(X) = -i X^0 X^1\;,
\end{equation}
In this model, there is a single complex scalar field $\tau$ that parametrizes the scalar manifold. To be precise, the above equation \eqref{eq:prepotential} may be expressed by $F(Z) = -i Z^0 Z^1$.

One can adopt a convenient gauge for projective coordinates by choosing
\begin{equation}
    Z^0 = 1, \qquad Z^1 = \tau.
\end{equation}
With this parametrization, the holomorphic symplectic vector becomes
\begin{equation}
    v(z) = 
    \begin{pmatrix}
        Z^0 \\ Z^1 \\ \partial_{Z^0} F(Z) \\ \partial_{Z^1} F(Z)
    \end{pmatrix}
    =
    \begin{pmatrix}
        1 \\ \tau \\ -i \tau \\ -i
    \end{pmatrix}\;.
\end{equation}
This vector encodes both the scalar field content (in the upper half) and its coupling to vector fields through the derivatives of the prepotential (in the lower half).

The Kähler potential is obtained by 
\begin{equation}
    e^{-K(\tau, \bar{\tau})} = -i \langle v(z), \bar{v}(\bar{z}) \rangle 
    = 2(\tau + \bar{\tau})\,,
\end{equation}
the Kähler metric for the scalar field is
\begin{equation}\label{eq:kahlermet}
    g_{\tau\bar{\tau}} = \partial_\tau \partial_{\bar{\tau}} K = \frac{1}{(\tau + \bar{\tau})^2}\,,
\end{equation}
and the period matrix $\mathcal{N}_{IJ}$ is calculated as
\begin{equation}\label{eq:period}
    \mathcal{N}_{IJ} = 
    \begin{pmatrix}
        -i \tau & 0 \\
        0 & -i/\tau
    \end{pmatrix}\;.
\end{equation}

This model is particularly appealing due to its simplicity while still capturing the key features of more general setups: non-trivial scalar dynamics, special Kähler geometry, and gauging via Fayet–Iliopoulos terms \cite{Bellucci:2008cb}. It serves as a prototypical example for the construction of explicit supersymmetric AdS$_4$ black holes, which we explore in detail in the next subsection.

Also, the scalar potential generated by the Fayet-Iliopoulos gauging is found to be
\begin{equation}
    V(\tau,\bar\tau) = -\frac{4}{\tau + \bar\tau} \left( g_0^2 + 2 g_0 g_1 (\tau + \bar\tau) + g_1^2 \tau \bar\tau \right)\;,
\end{equation}
which admits an AdS vacuum when $\tau = \bar\tau = |g_0/g_1|$, provided that $g_0 g_1 > 0$ to ensure a real and positive solution.
\subsection*{Real scalar truncation}
In this paper, we consider the simple real scalar truncation of the above theory by demanding $\tau = \bar\tau$. This is a consistent truncation as the resulting theory includes the BPS black hole solution found in \cite{Cacciatori:2009iz}.  With this truncation, the real part of the period matrix vanishes $Re \mathcal{N}_{IJ}=0$, and the scalar potential becomes
\begin{equation}\label{eq:scalarpot}
    V(\tau)=\frac{-2}{\tau}(g^2_0+4 g_0 g_1 \tau+g^2_1\tau^2)\,,
\end{equation}
Then, the bosonic Lagrangian \eqref{eq:generalL}  becomes
\begin{equation}\label{eq:lagrangian}
\begin{split}
    \mathcal{L}&=-\frac{\tau}{4}  F^{(0)\mu\nu} F^{(0)}_{\mu\nu}-\frac{1}{4\tau} F^{(1)\mu\nu} F^{(1)}_{\mu\nu}+\frac{R}{16 \pi G_N} -\frac{1}{4\tau^2}\partial_\mu \tau \partial^\mu \tau -V(\tau)\,.
\end{split}
\end{equation}
This is nothing but Einstein-Dilaton-Maxwell theory with a scalar potential, where we shall naturally parametrize the scalar by the dilaton as $\tau=e^{-2\Phi}$.

\subsection{AdS$_4$ BPS black hole solution and the near horizon}\label{app:AdS4blackhole}
A general class of timelike supersymmetric backgrounds was classified in \cite{Cacciatori:2008ek}.  
In a set of adapted coordinates, the metric can be written as
\begin{equation}\label{eq:genmetric}
    ds^2 = -4 |b|^2 (dt + \sigma)^2 + |b|^{-2} \left( dz^2 + e^{2\Phi} \, dw \, d\bar{w} \right),
\end{equation}
where $b(z,w,\bar{w})$ and $\Phi(z,w,\bar{w})$ are functions of the coordinates.
The two-dimensional metric $e^{2\Phi} dw d\bar{w}$ describes a Riemann surface $\Sigma_\kappa$ of constant curvature $\kappa = 0, \pm 1$.

Restricting to static configurations with $\sigma = 0$ in \eqref{eq:genmetric}, fields depending only on the radial coordinate $z$ lead to black hole solutions, carrying electric and magnetic charges $(q_I, p^I)$, which satisfy the Dirac quantisation conditions.
For the class of solutions of interest, we take $\kappa = -1$, corresponding to a hyperbolic horizon geometry \cite{Cacciatori:2009iz}.

Considering  real scalar fields, setting $\tau = \bar{\tau}$, and assuming $b(z)$ in \eqref{eq:genmetric}, to be purely imaginary, one obtains,
\begin{equation}
    b(z) = i N(z), \qquad N(z) \in \mathbb{R}.
\end{equation}
This choice simplifies the BPS equations while ensuring a real lapse function $N(z)$.  
Under these assumptions, the purely magnetic configuration is characterised by 
\begin{equation}
    q_I = 0, \qquad p^I \neq 0\;,
\end{equation}
where $p^I$ denotes the magnetic charges.
The magnetic charges $ p^I $ are subject to the Dirac quantisation condition $\kappa = -8\pi g_I p^I = -1$, Furthermore, the charges are also constrained by the relation  
\begin{equation}
    g_0 p^0 = g_1 p^1\;.
\end{equation}
From this, we obtain the relation   
\begin{equation}\label{eq:diracquant}
    p^I = \frac{1}{16\pi g_I}\;,
\end{equation}
indicating that the magnetic charges are quantised inversely with respect to the corresponding coupling parameters $ g_I$.  
\medskip
 Introducing a parameter \(\nu\) via the relation \(\sinh\nu = 2\sqrt{2} g_0 \beta^0\), the metric simplifies to
\begin{equation}\label{eq:fullmetric}
    ds^2 = -4 N^2(z)\, dt^2 + \frac{dz^2}{N^2(z)} + \frac{z^2 - \tfrac{1}{2} \sinh^2 \nu}{4 g_0 g_1}\, e^{2\gamma} dw d\bar{w}\;,
\end{equation}
with the lapse function $N^2(z)$ given explicitly by
\begin{equation}\label{eq:lapsefunc}
    N^2(z) = \frac{4 g_0 g_1 \left(z^2 - \frac{1}{2}\cosh^2 \nu \right)^2}{z^2 - \tfrac{1}{2} \sinh^2 \nu}\;.
\end{equation}
The location of the event horizon can be determined from the condition 
$N^2(z_h) = 0$, which yields
\begin{equation}\label{eq:horizonloc}
z_h = \frac{1}{\sqrt{2}}\,\cosh\nu.
\end{equation}
The scalar field and the magnetic field strengths take the form
\begin{equation}\label{eq:fieldsolution}
    \tau(z) = \frac{g_0}{g_1} \cdot \frac{z - \tfrac{1}{\sqrt{2}} \sinh \nu}{z + \tfrac{1}{\sqrt{2}} \sinh \nu}\, \qquad
    F^I = 2\pi i \, p^I \, e^{2\gamma} \, dw \wedge d\bar{w}\;.
\end{equation}


\subsection*{ Near-horizon geometry}
The near horizon geometry AdS$_2 \times \mathbb{H}^2$ is obtained by expanding the metric around the horizon  $z_h = \frac{1}{\sqrt{2}} \cosh\nu$. Redefining the coordinate as
\begin{equation}\label{NearhorizonLimit}
    z \rightarrow  z_h +\epsilon {r}\;, \,\,\,\, t\rightarrow 2L_1^2 \frac{\theta}{\epsilon}\,,
\end{equation}
and subsequently taking the scaling limit  $\epsilon \rightarrow 0$ corresponds to the near-horizon limit.\footnote{Note that the limit $\epsilon \rightarrow 0$ in \eqref{NearhorizonLimit} is being taken after the extremal limit, and it would result in 
\begin{equation}
     ds^2= L_1^2  \left({r}^2 d {\theta}^2+\frac{d {r}^2}{{r}^2 }\right)\nonumber\,.
\end{equation}
To obtain \eqref{eq:nearhormetric}, we assume to start with a non-extremal solution and take the extremal and near-horizon limit simultaneously. 
}
The resulting near-horizon geometry is given by
\begin{equation}\label{eq:nearhormetric}
    ds^2= L_1^2 \bigg( ({r}^2-1) d \theta^2+\frac{d {r}^2}{({r}^2-1) }\bigg)+L_2^2\left( d\psi^2 + \sinh^2\psi\, d\vartheta^2 \right)\;,
\end{equation}
with 
 $L_1^2=(16  g_0 g_1 \cosh^2\nu )^{-1} \,\quad{\rm and}\quad L_2^2=(8 g_0 g_1)^{-1} $, where we have rewritten the metric of the two-dimensional hypersurface in terms of coordinates $\psi,\vartheta$. The near-horizon configurations of fields and their properties are presented in Section \ref{sec:BH background}.

\paragraph{Volume of AdS$_2$ Part}

Considering Euclidean AdS$_2$ of radius $L_1$ 
\begin{equation}\label{eq:adsmetric}
ds^2 = L_1^2 \bigg( ({r}^2-1) d \theta^2+\frac{d {r}^2}{({r}^2-1) }\bigg)\;, 
\end{equation}

\begin{equation}
\sqrt{ g}_{\text{AdS}_2} = L_1^2\;.
\end{equation}
$( r,\theta)$ represent coordinates describing AdS$_2$ within the range $1 \leq r\leq \infty$ and $0\leq\theta < 2\pi$. We can introduce an infrared cut off at $r =\tilde{r}_c$
\begin{equation}
\text{Vol} 
= \int_0^{2\pi} d\theta \int_{1}^{\tilde{r}_c} dr \,\sqrt{g_{\text{AdS}_2}}
= 2\pi L_1^2 \,(\tilde{r}_c-1)\;,
\end{equation}
which diverges linearly as $\tilde{r}_c\to\infty$.
 The finite part of the integral is Vol$=- 2\pi L^2_1$\;.
The divergent term proportional to $r_c$ cancels, leaving the universal finite result
\begin{equation}
{\;\text{Vol}_{\text{AdS}_2} = - 2\pi L^2_1 \;.}
\end{equation}

\paragraph{Volume of H$_2$ Part}
In the case of the hyperbolic part, volume is infinite, thus we express the volume as,
\begin{equation}
{\text{Vol}_{\mathbb{H}^2} = L^2_2 \mathbb{V}_{\mathbb{H}^2} }\; ,
\end{equation}
 where, $ \mathbb{V}_{\mathbb{H}^2}$ is the volume of space with unit radius.

Thus, the renormalised volume for the near-horizon geometry is
\begin{equation}
\text{Vol}_{\text{ren}} 
= (-2\pi L^2_1 L_2^2) \;\mathbb{V}_{\mathbb{H}^2} .
\end{equation}

\section{Useful definitions and identities}\label{app:Identities}
{\bf Wheeler-DeWitt metric:}
\begin{equation}
    {I}_{ab,cd} = \frac{1}{2}\left(g_{ac}g_{bd}+g_{ad}g_{bc}- g_{ab}g_{cd}\right)\,.
\end{equation}
The inverse is given by 
\begin{equation}
    {I}^{ab,cd} = \frac{1}{2}\left(g^{ac}g^{bd}+g^{ad}g^{bc}- g^{ab}g^{cd}\right)\,,
\end{equation}
such that 
\begin{equation}\label{eq:InverseRelation}
     {I}_{ab,ef}  {I}^{ef,cd} \;=\; \mathrm{1}_{ab}^{\;cd} \;\equiv\; \delta_{(a}^{\;c} \delta_{b)}^{\;d} \,.
\end{equation}
We shall denote ${I}_{ab}{}^{cd}$ as the one whose indices are raised by the spacetime metric $g^{ab}$
\begin{equation}
    {I}_{ab}{}^{cd} \equiv {I}_{ab, ef}\,  g^{ce} g^{df}\,.
\end{equation}

Since the square of the Wheeler-DeWitt metric gives the identity as \eqref{eq:InverseRelation}, one can naturally define the projector as
\begin{equation}
    (P^{\text{TL}})_{ab}{}^{cd} = \frac{1}{2}\left( \mathrm{1} + {I}\right)_{ab}{^{cd}}\,,\qquad (P^{\text{tr}})=\frac{1}{2}\left( \mathrm{1} - {I}\right)_{ab}{^{cd}}\,,
\end{equation}
where $P^{\text{TL}}$ and $P^{\text{tr}}$ denote the projection onto the traceless part and trace part, respectively. The trace projector is 
\begin{equation}
    (P^{\text{tr}})_{ab}{}^{cd} = \frac{1}{4}g_{ab}g^{cd}\,,
\end{equation}
and the others are related by
\begin{equation}
    \mathrm{1}_{ab}{}^{cd} =  (P^{\text{TL}})_{ab}{}^{cd}+(P^{\text{tr}})_{ab}{}^{cd}\,,
\end{equation}
\begin{equation}
    {I}_{ab}{}^{cd} =  (P^{\text{TL}})_{ab}{}^{cd}-(P^{\text{tr}})_{ab}{}^{cd}\,.
\end{equation}
Note that the Wheeler-DeWitt metric naturally defines the inner product for gravitons.  Because of the negative sign of the trace projection part, the DeWitt metric has an indefinite sign, and it represents the wrong sign of the conformal mode of the graviton.\\

{\bf Lichnerowicz operator}: the kinetic operator that naturally appears for  the graviton mode~$h_{ab}$ defined as
\begin{equation}
 (   \Delta_{\text{L}})_{ab}{}^{cd} = - \mathrm{1}_{ab}^{\;cd}\nabla^2 + ({M}_{\text{L}})_{ab}{}^{cd}\,, 
\end{equation}
where we have denoted the mass-like term as
\begin{equation}
    (M_{\text{L}})_{ab}{}^{cd} \equiv -R_{a}{}^{c}{}_{b}{}^{d}-R_{b}{}^{c}{}_{a}{}^{d}+ R_{a}{}^{(c}\delta_{b}^{\;d)}+ R_{b}{}^{(c}\delta_{a}^{\;d)}\,.
\end{equation}
Note that the mass term is the kernel of the trace-projector, i.e.
\begin{equation}
(    P^{\text{tr}})_{ab}{}^{ef}(M_{\text{L}})_{ef}{}^{cd}=0\,.
\end{equation}
Therefore, the action of the DeWitt metric on the mass term is trivial
\begin{equation}
        {I}_{ab}{}^{ef}(M_{\text{L}})_{ef}{}^{cd}=  (    \mathrm{1}-2P^{\text{tr}} )_{ab}{}^{ef}(M_{\text{L}})_{ef}{}^{cd}= (M_{\text{L}})_{ab}{}^{cd}\,.
\end{equation}
Note that the trace of this mass term is
\begin{equation}
    (M_{\text{L}})_{ab}{}^{ab}= 6R\,.
\end{equation}

{\bf Cayley-Hamilton identity}: for antisymmetric $4\times 4$ tensors $F$ and $G$, we have the following identity,
\begin{equation}\label{CH-identity1}
    {\rm Tr}(F^2 G^2 ) = \frac{1}{4}\left[({\rm Tr}(F^2))({\rm Tr}(G^2))+ ({\rm Tr}(FG))^2 - ({\rm Tr}(F \widetilde{G}))^2\right]\,,
\end{equation}
where the dual tensor is denoted by $\widetilde{G}_{ab} = \frac{1}{2}\, \epsilon_{abcd}G^{cd}$. A particular case when $F= G$, the identity is reduced to
\begin{equation}\label{CH-identity2}
    {\rm Tr}(F^4) = \frac{1}{2}({\rm Tr}(F^2))^2 - \frac{1}{4}({\rm Tr}(F \widetilde{F}))^2\,.
\end{equation}
\subsection*{Background Identities}\label{app:backidentites}

For the on-shell background fields, we have the following identities.
\begin{itemize}
\item 
Using the Maxwell equation and the Bianchi identity, up to the total derivatives, the Maxwell fields satisfy the following relations
\begin{align}\label{eq:rule1}
    \begin{split}
        (D_\rho \bar{F}^{(I)}_{\mu\nu})
    (D^\rho \bar{F}_{}^{(I)\mu\nu})
    &= 2(D_\rho \bar{F}^{(I)}_{\mu\nu})
    (D^\nu \bar{F}^{(I)\mu\rho})\\&
    =\bar R_{\mu\nu\rho\sigma}\,
       \bar{F}_{}^{(I)\mu\nu}
       \bar{F}_{}^{(I)\rho\sigma}
       -2\bar R_{\mu\nu}\,
        \bar{F}_{}^{\mu(I)}{}_{\rho}
        \bar{F}_{}^{(I)\nu\rho}\,.
    \end{split}
\end{align}
From symmetry properties of the Riemann tensor, we can use the following identities
\begin{equation}\label{identity:Rieman-F}
    \bar R_{\nu\sigma\mu\rho}\,
    \bar{F}^{\mu\nu}\bar{F}^{\rho\sigma}
    =\frac{1}{2}\,
    \bar R_{\mu\nu\rho\sigma}\,
    \bar{F}^{\mu\nu}\bar{F}^{\rho\sigma},
    \quad
    \bar R_{\mu[\nu\rho\sigma]}=0,\quad \bar R_{\mu\rho\nu\sigma}\bar R^{\mu\nu\rho\sigma}=\frac{1}{2} \bar R_{\mu\nu\rho\sigma}\bar R^{\mu\nu\rho\sigma}\,.
\end{equation}

\item Note that our background is purely magnetic such that $\bar{F}^{(I)}\widetilde{\bar{F}}^{(J)}=0$, and by the ansatz the two field strength are aligned along the same horzion two form so that $\bar F^{(0)}\propto \bar F^{(1)}$. With those two properties, the Cayley--Hamilton identities \eqref{CH-identity1} and
\eqref{CH-identity2} obtains the following identities: 
\begin{eqnarray}
\label{CH-identity_F1=F2}
&&\bar{F}^{(0)}_{\mu\rho}\bar{F}_{\;\nu}^{(1)\rho}\bar{F}^{(0)\mu\sigma}\bar{F}^{(1)\nu}{}_{\sigma} = \frac{1}{2} (\bar{F}^{(0)})^2(\bar{F}^{(1)})^2  =\frac{1}{2}(\bar{F}^{(0)}\bar{F}^{(1)})^2\;,
\\
\label{CH-identity_F12}
    &&  \bar{F}^{(0)}_{\mu\rho}\bar{F}_{\;\nu}^{(0)\rho}\bar{F}^{(0)\mu\sigma}\bar{F}^{(0)\nu}{}_{\sigma} = \frac{1}{2} (\bar{F}^{(0)})^2(\bar{F}^{(0)})^2\;,
    \\
    &&  \bar{F}^{(1)}_{\mu\rho}\bar{F}_{\;\nu}^{(1)\rho}\bar{F}^{(1)\mu\sigma}\bar{F}^{(1)\nu}{}_{\sigma} = \frac{1}{2} (\bar{F}^{(1)})^2(\bar{F}^{(1)})^2\;.
\end{eqnarray}

\item 
These relations between the two gauge fields combine with the graviton equation of motion to obtain further identities: by taking the square of the left-hand side and right-hand side of the \eqref{eq:gravitoneom} we have

 \begin{equation}\label{ruleF2=R2}
       \Bigl(e^{-2\phi_0}(\bar{F}^{(0)})^2 + e^{2\phi_0 }(\bar{F}^{(1)})^2\Bigr)^2 = \bar R_{\mu\nu}\bar R^{\mu\nu}-\frac{1}{4}\bar R^2\,.
   \end{equation}
Moreover, by multiplying $R^{\mu\nu}$ to the both side of \eqref{eq:gravitoneom}, we have 
\begin{align}\label{rule:RF-R2}
    \begin{split}
    &\bar R^{\mu\nu}\left(e^{-2\phi_0 }\bar F_{\mu\rho}^{(0)}\bar F_{\;\nu}^{(0)\rho}+e^{2\phi_0 }\bar F_{\mu\rho}^{(1)}\bar F_{\;\nu}^{(1)\rho}\right)   
    \\
   & \;\; = \frac{1}{2}\bar R_{\mu\nu}\bar R^{\mu\nu} -\frac{1}{8}\bar R^2+ \frac{1}{4} \bar R \left(e^{-2\phi_0 }(\bar F^{(0)})^2+e^{2\phi_0 }(\bar F^{(1)})^2\right)  \,.
    \end{split}
\end{align}

\end{itemize}

\section{Detailed Calculations}\label{app:detailedcal}
 Note that the index $m$ is raised and lowered by the metric $I^{mn}$ and $I_{mn}$, respectively.  Therefore, it defines the multiplication of two operators $A^{mn}$ and $B^{mn}$ as
\begin{eqnarray}\label{rule_product}
    (A\cdot B)^{mn} = A^{mp}B_{p}{}^{n} \equiv A^{mp}I_{pq}B^{qn}\,, 
\end{eqnarray}
and also the trace is defined as 
\begin{equation}\label{rule_trace}
    {\rm Tr} (A) \equiv I_{mn}A^{nm}\,.
\end{equation}
With this rule, we present the explicit evaluation of the field-space operators $ (\omega^\rho \omega_\rho)$, $\quad [D_\rho, D_\sigma]$ , $(D_{[\rho }\omega_{\sigma]})$  and $[\omega_\rho,\omega_\sigma]$ required for $E$ and $\Omega_{\rho\sigma}$ and their respective trace, which enter the Seeley-DeWitt coefficient $a_4$.  

\subsection*{For $E_m{}^n$ 
}

\begin{eqnarray}
    \xi_m \left(D_\rho \omega^\rho\right)^{mn} \xi_n&=&- \sqrt{2}e^{-\phi_0} a^{(0)}_\sigma  D^{(\mu}\bar F^{(0)\nu)\sigma} h_{\mu\nu}+\sqrt{2}e^{-\phi_0} h_{\mu\nu} D^{(\mu}\bar F^{(0)\nu)\sigma}a^{(0)}_\sigma\nonumber\\&&
    - \sqrt{2}e^{\phi_0} a^{(1)}_\sigma D^{(\mu}\bar F^{(1)\nu)\sigma} h_{\mu\nu}+\sqrt{2}e^{\phi_0} h_{\mu\nu} D^{(\mu}\bar F^{(1)\nu)\sigma}a^{(1)}_\sigma\,.
\end{eqnarray}

Note that $\omega_\rho$ is traceless with respect to the spacetime metric $g_{\mu\nu}$
\begin{equation}
    g_{\mu\nu} (\omega_{\rho})^{\mu\nu\,  m}=0\,.
\end{equation}
Therefore, all the spacetime indices in $(\omega_\rho)^{mn}$ can be raised and lowered by the spacetime metric.  The multiplication $(\omega_\rho  \omega^\rho)^{mn}= (\omega_\rho)^{mp}  I_{pq}(\omega^\rho)^{qn}$ can be done only with spacetime metric. 
\begin{eqnarray}
    \xi_m (\omega^\rho \omega_\rho)^{mn} \xi_n \! &=& h^{\mu\nu }\!\left\{-2 \delta_{(\mu}^{(\lambda} \bar{R}_{\nu)}{}^{\sigma)} +\bar{g}_{\mu\nu}\bar{R}^{\lambda \sigma} + \bar{R}_{\mu\nu}\bar{g}^{\lambda\sigma} +\tfrac{1}{2} (\delta_{(\mu}^{\lambda}\delta_{\nu)}^{\sigma} - g_{\mu\nu}g^{\lambda\sigma})\bar{R} 
    \right.\nonumber\\
    &&\left.
\qquad
-\bigl(\delta_{(\mu}^{\lambda}\delta_{\nu)}^{\sigma} - \tfrac{1}{2}\bar{g}_{\mu\nu}\bar{g}^{\lambda\sigma}\bigr)\bigl( e^{-2\phi_0}(\bar{F}^{(0) } )^2+ e^{2\phi_0}(\bar{F}^{(1)})^2 \bigr) \right.\nonumber
    \\
    &&\left. \qquad-4 \,e^{-2\phi_0}\bar{F}^{(0)}_{{(\mu}}{}^{\lambda}\bar{F}^{(0)}_{{\nu)}}{}^{\sigma}-4 \,e^{2\phi_0}\bar{F}^{(1)}_{{(\mu}}{}^{\lambda}\bar{F}^{(1)}_{{\nu)}}{}^{\sigma} \right\}h_{\lambda \sigma}\nonumber
    \\
    &&-a^{(0)}_{\mu}e^{-2\phi_0}\left( \bar{g}^{\mu\nu}(\bar{F}^{(0)})^2 +3 \bar{F}_{\;\;\rho}^{(0)\mu}\bar{F}^{(0)\rho\nu}\right)a^{(0)}_{\nu} \nonumber
    \\
     &&-a^{(1)}_{\mu}e^{2\phi_0}\left( \bar{g}^{\mu\nu}(\bar{F}^{(1)})^2 +3 \bar{F}_{\;\;\rho}^{(1)\mu}\bar{F}^{(1)\rho\nu}\right)a^{(1)}_{\nu} 
     \\
     &&- \phi \left( e^{-2\phi_0}(\bar{F}^{(0)})^2 + e^{2\phi_0}(\bar{F}^{(1)})^2  \right)\phi \nonumber
     \\
     && - a^{(0)}_{\mu }\left(\bar{g}^{\mu\nu}\bar{F}^{(0)}_{\rho\sigma} \bar{F}^{(1)\rho\sigma} + \bar{F}_{\;\;\rho}^{(1)\mu} \bar{F}^{(0)\rho \nu}\right)a^{(1)}_{\nu} \nonumber
     \\
     && - a^{(1)}_{\mu }\left(\bar{g}^{\mu\nu}F^{(0)}_{\rho\sigma} \bar{F}^{(1)\rho\sigma} + \bar{F}_{\;\;\rho}^{(0)\mu}\bar{F}^{(1)\rho \nu}\right)a^{(0)}_{\nu} \nonumber
     \\
     && +(h_{\mu\nu}\phi +\phi h_{\mu\nu})\tfrac{1}{\sqrt{2}}\left\{ e^{-2\phi_0}\bigl(\bar{g}^{\mu\nu}(\bar{F}^{(0)})^2 - 4 \bar{F}_{\;\;\rho}^{(0)\mu}\bar{F}^{(0)\rho\nu} )\bigr)\right. \nonumber
     \\
     &&\qquad\qquad\qquad\qquad\qquad \left. -e^{2\phi_0}\bigl(\bar{g}^{\mu\nu}(\bar{F}^{(1)})^2 - 4 \bar{F}_{\;\;\rho}^{(1)\mu}\bar{F}^{(1)\rho\nu} )\bigr) \right\}\,.\nonumber
\end{eqnarray}

\begin{eqnarray}
    \xi^m E_{m}{}^{n}\xi_n &=& h^{\mu\nu}\left( 2 R_{(\mu}{}^{\lambda}{}_{\nu)}{}^{\sigma} - R_{\mu\nu}g^{\lambda\sigma} +\tfrac{1}{4}g_{\mu\nu}g^{\lambda\sigma}R  \right)h_{\lambda \sigma}\nonumber
    \\
    &&+a^{(0)\mu} \left\{ e^{-2\phi_0}\left(3\bar F_{\;\mu\rho}^{(0)}\bar F^{(0)\nu\rho}+\delta_\mu^{\;\nu}(\bar F^{(0)})^2 \right)-\bar R_{\mu}{}^{\nu}\right\}a^{(0)}_\nu
     \nonumber\\
     &&+a^{(1)\mu} \left\{ e^{2\phi_0}\left(3\bar F_{\;\mu\rho}^{(1)}\bar F^{(1)\nu\rho}+\delta_\mu^{\;\nu}(\bar F^{(1)})^2 \right)-\bar R_{\mu}{}^{\nu}\right\}a^{(1)}_\nu
     \nonumber\\
      &&+ \phi \left( e^{-2\phi_0}(\bar{F}^{(0)})^2 + e^{2\phi_0}(\bar{F}^{(1)})^2  \right)\phi \nonumber
      \\
      && -h^{\mu \nu} \sqrt{2} e^{-\phi_0} D_{(\mu}F_{\nu)}^{(0)\lambda} a^{(0)}_{\lambda} -a^{(0)\lambda}\sqrt{2} e^{-\phi_0} D^{(\mu}F^{(0)\nu)}{}_{\lambda} h_{\mu\nu} \nonumber
          \\
      && -h^{\mu \nu} \sqrt{2} e^{\phi_0} D_{(\mu}F_{\nu)}^{(1)\lambda} a^{(1)}_{\lambda} -a^{(1)\lambda}\sqrt{2} e^{\phi_0} D^{(\mu}F^{(1)\nu)}{}_{\lambda} h_{\mu\nu} \,\nonumber
       \\
     && + a^{(0)\mu}\left(\delta_\mu^{\nu}\bar{F}^{(0)}_{\rho\sigma} \bar{F}^{(1)\rho\sigma} + \bar{F}_{\;\rho\mu}^{(1)} \bar{F}^{(0)\rho \nu}\right)a^{(1)}_{\nu} \nonumber
     \\
     && +a^{(1)\mu}\left(\delta_{\mu}^{\nu}F^{(0)}_{\rho\sigma} \bar{F}^{(1)\rho\sigma} + \bar{F}_{\;\rho\mu}^{(0)}\bar{F}^{(1)\rho \nu}\right)a^{(0)}_{\nu} \,\nonumber\\&&+h^{\mu\nu} \frac{1}{\sqrt{2}}g_{\mu\nu} \left(e^{-2\phi_0}(F^{(0)})^2-e^{2\phi_0}(F^{(1)})^2\right) \phi
\nonumber\\&&
+ \phi\, 
\frac{1}{\sqrt{2}}{g^{\mu\nu}}\left(-e^{-2\phi_0}(F^{(0)})^2+e^{2\phi_0}(F^{(1)})^2\right)h_{\mu\nu}\,.
\end{eqnarray}
The trace gives
\begin{equation}
  {\rm Tr}(\omega_\rho \omega^\rho)=   (\omega^\rho \omega_\rho)_m{}^{m} = -14 \left(e^{-2\phi_0}(F^{(0)})^2+ e^{2\phi_0}(F^{(1)})^2 \right)\,.
\end{equation}

\begin{equation}
    {\rm Tr}( P) \equiv (\mathbb{I}P)_m{}^{m}= -3R -6 \left(e^{-2\phi_0}(F^{(0)})^2 + e^{2\phi_0} (F^{(1)})^2 \right)\,. 
\end{equation}

\subsection*{For ${\rm Tr}E^2$ 
}
Let us denote $E_m{}^n$ in terms of $4\times 4$ block matrices such that each block acts on $\{h_{\mu\nu}, a^{(0)}_\rho , a^{(1)}_\sigma , \phi \}$ fields space. That is, we are denoting
\begin{equation}
    E_m{}^n = \begin{pmatrix}
        E_{hh} &E_{h0}& E_{h1}& E_{h\phi}\\
        E_{0h}&E_{00} &E_{01}& E_{0\phi}\\
        E_{1h}&E_{10} &E_{11}& E_{1\phi}\\
        E_{\phi h}&E_{\phi 0} &E_{\phi1}& E_{\phi\phi}        
    \end{pmatrix}\,, \qquad (E^2)_m{}^n =\begin{pmatrix}
        E^2_{hh} &E^2_{h0}& E^2_{h1}& E^2_{h\phi}\\
        E^2_{0h}&E^2_{00} &E^2_{01}& E^2_{0\phi}\\
        E^2_{1h}&E^2_{10} &E^2_{11}& E^2_{1\phi}\\
        E^2_{\phi h}&E^2_{\phi 0} &E^2_{\phi1}& E^2_{\phi\phi}   
    \end{pmatrix}\,,
\end{equation}
where $E_{0\phi}= E_{1\phi}= E_{\phi 0 } =E_{\phi 1}=0$.\\
We want to compute 
\begin{align}
    \begin{split}
         {\rm Tr} E^2 &= {\rm tr}E^2_{hh} \,+{\rm tr}E^2_{00}\,+{\rm tr}E^2_{11}\, +{\rm tr}E^2_{\phi\phi}
         \\
 &=       {\rm tr}(E_{hh})^2 \,+{\rm tr}(E_{00})^2\,+{\rm tr}(E_{11})^2\, +{\rm tr}(E_{\phi\phi})^2\\
 &\quad +2 {\rm tr}\big( E_{h0}E_{0h }\, +E_{h1}E_{1h }\, + E_{01}E_{10 }\, +E_{h\phi}E_{\phi h }\bigr)\,.
    \end{split}
\end{align}
where the trace with small letter denotes the trace in each block. 

The trace of each term is obtained, using the identities \eqref{app:backidentites}, as follows:
\begin{eqnarray}
     {\rm tr} (E_{hh})^2 &=& 3\bar R_{\mu\nu\rho\sigma}\bar R^{\mu\nu\rho\sigma}-4\bar R_{\mu\nu}\bar R^{\mu\nu}+\bar R^2\,,\nonumber
     \\
     {\rm tr} (E_{00})^2 &=&\!\frac{29}{2}e^{-4\phi_0} (\bar{F}^{(0)})^2 (\bar{F}^{(0)})^2  \!-\! e^{-2\phi_0}\!\left( \!6 \bar{R}^{\mu\nu}\bar{F}^{(0)}_{\mu \rho}\bar{F}_{\;\nu}^{(0)\rho} +2 \bar{R}(\bar{F}^{(0)})^2  \right)\!+ \!\bar R_{\mu\nu}\bar R^{\mu\nu}\,,\nonumber
      \\
      {\rm tr} (E_{11})^2 &=&\frac{29}{2}e^{4\phi_0} (\bar{F}^{(1)})^2 (\bar{F}^{(1)})^2  - e^{2\phi_0}\left( 6 \bar{R}^{\mu\nu}\bar{F}^{(1)}_{\mu \rho}\bar{F}_{\;\nu}^{(1)\rho} +2 \bar{R}(\bar{F}^{(1)})^2  \right)+ \bar R_{\mu\nu}\bar R^{\mu\nu}\,,\nonumber
      \\
       {\rm tr} (E_{\phi\phi})^2 &=& \bar R_{\mu\nu}\bar R^{\mu\nu} -\frac{1}{4}\bar R^2\,,\nonumber
       \\
       2{\rm tr} (E_{h0}E_{0h}) &=& e^{-2\phi_0}\left( 3\bar{R}^{\mu\nu \rho \sigma}\bar{F}^{(0)}_{\mu\nu}\bar{F}^{(0)}_{\rho\sigma} -6 \bar R^{\mu\nu} \bar{F}_{\mu\rho}^{(0)}\bar{F}_{\;\nu}^{(0)\rho}\right)\,,
       \\
       2{\rm tr} (E_{h1}E_{1h}) &=& e^{2\phi_0}\left( 3\bar{R}^{\mu\nu \rho \sigma}\bar{F}^{(1)}_{\mu\nu}\bar{F}^{(1)}_{\rho\sigma} -6 \bar R^{\mu\nu} \bar{F}_{\mu\rho}^{(1)}\bar{F}_{\;\nu}^{(1)\rho}\right)\,,\nonumber
       \\
        2{\rm tr} (E_{01}E_{10}) &=& 13 (\bar{F}^{(0)})^2(\bar{F}^{(1)})^2\,,\nonumber
        \\
         2{\rm tr} (E_{h\phi}E_{\phi h}) &=& -4\left(e^{-2\phi_0} (\bar{F}^{(0)})^2 -e^{2\phi_0} (\bar{F}^{(1)})^2 \right)^2\,.\nonumber 
\end{eqnarray}
Summing over all the terms above and using the background identities in \ref{app:Identities}, we obtain
\begin{eqnarray}\label{app:traceE2}
    \text{Tr} (E^2)&=& 3 \bar R_{\mu\nu\rho\sigma}\bar R^{\mu\nu\rho\sigma}+3\left(e^{-2\phi_0} \bar F^{(0)\mu\nu}\bar F^{(0)\lambda\sigma}+e^{2\phi_0} \bar F^{(1)\mu\nu}\bar F^{(1)\lambda\sigma}\right)\bar R_{\mu\nu\lambda\sigma}\nonumber\\
    &&+\frac{7}{2}\bar R_{\mu\nu} \bar R^{\mu\nu}-\frac{3}{8}\bar R^2-5 \bar R(e^{-2\phi_0}(\bar F^{(0)})^2+e^{2\phi_0}(\bar F^{(1)})^2)\,.
\end{eqnarray}
\subsection*{For  $\Omega_{\rho\sigma}$
}
Next, we proceed to evaluate the commutator of covariant derivatives, 
$[D_\rho, D_\sigma]$, acting on the different fluctuation fields. 
It is important to note that while the covariant derivatives commute when acting on scalar quantities (such as the scalar fluctuation), they do not commute for fields—such as vectors, tensors, or spinors—where curvature and gauge-field strengths contribute nontrivially through their respective commutator relations. 
\begin{equation}
    \begin{split}
       \xi_m [D_\rho, D_\sigma]^{mn}\xi_n &=\frac{1}{2}h_{\mu\nu}( \bar g^{\mu\lambda}\bar R^{\nu\kappa}{}_{\rho\sigma}+\bar g^{\nu\lambda}\bar R^{\mu\kappa}{}_{\rho\sigma}+\bar g^{\mu\kappa}\bar R^{\nu\lambda}{}
       _{\rho\sigma}+\bar g^{\nu\kappa}\bar R^{\mu\lambda}{}_{\rho\sigma})h_{\lambda\kappa}\\
       &\quad + \,a^{(0)}_\mu \bar R^{\mu\nu}{}_{\rho\sigma}a^{(0)}_\nu+a^{(1)}_\mu\bar R^{\mu\nu}{}_{\rho\sigma}a^{(1)}_\nu \,.
    \end{split}
\end{equation}
Note that $[
D_\rho\,,D_\sigma
]_m {}^n$ whose index  is lowered by the metric $I_{mn}$ is the same as being lowered by the spacetime metric $g_{\mu\nu}$. 
\\

The commutation between the two connections $\omega_\rho$ and $\omega_\sigma$ is given by,
\begin{eqnarray}
\xi_m [\omega_\rho,\omega_\sigma]^{mn} \xi_n
&=& \xi_m \big(\omega_\rho \omega_\sigma- \omega_\sigma \omega_\rho\big)^{mn} \xi_n
\nonumber\\[4pt]
&=& a^{(0)}_{\mu}\, e^{-2\phi_0}
\Big\{-6\bar F^{\mu(0)}
_{\;\;[\rho}\bar F^{(0)\nu}
_{\sigma]}+4\delta^{[\nu}_{\,\,[\rho}  \bar F^{(0)}_{\sigma]\lambda}\bar F^{\mu]\lambda(0)}-4\bar F^{(0)\mu\nu}\bar F^{(0)}_{\rho\sigma}\Big\}
a^{(0)}_{\nu} \nonumber\\&& + a^{(1)}_{\mu}\, e^{2\phi_0}\,
\Big\{
-6\bar F^{\mu(1)}
_{\;\;[\rho}\bar F^{(1)\nu}
_{\sigma]}+4\delta^{[\nu}_{\,\,[\rho}  \bar F^{(1)}_{\sigma]\lambda}\bar F^{\mu]\lambda(1)}-4\bar F^{(1)\mu\nu}\bar F^{(1)}_{\rho\sigma}
\Big\}
a^{(1)}_{\nu}\nonumber\\&&+a^{(0)}_{\mu}\Big\{-2\bar F^{(0)}_{\rho\sigma} \bar F^{(1)\mu\nu}-2\bar F^{(1)}_{\rho\sigma} \bar F^{(0)\mu\nu}+2\bar F^{\nu(0)}_{\,\,[\rho} \bar F^{(1)\mu}_{\sigma]}-8 \bar F^{\mu(0)}_{\;\;[\rho}\bar F^{(1)\nu}_{\sigma]}\nonumber\\&&\qquad\qquad+2\delta^{\nu}_{\;\;[\rho}  \bar F^{(1)}_{\sigma]\lambda}\bar F^{(0)\mu\lambda}-2\delta^{\mu}_{\;\;[\rho}  \bar F^{(0)}_{\sigma]\lambda}\bar F^{(1)\nu\lambda}+2 g^{\mu\nu} \bar F^{\lambda(0)}_{\;\;\;\;[\rho}\bar F^{(1)}_{\sigma]\lambda}\Big\}a^{(1)}_{\nu}\nonumber\\&&+a^{(1)}_{\mu}\Big\{-2\bar F^{(0)}_{\rho\sigma} \bar F^{(1)\mu\nu}-2\bar F^{(1)}_{\rho\sigma} \bar F^{(0)\mu\nu}-2\bar F^{\mu(0)}_{\,\,[\rho} \bar F^{(1)\nu}_{\sigma]}+8 \bar F^{\nu(0)}_{\;\;[\rho}\bar F^{(1)\mu]}_{\sigma]}\nonumber\\&&\qquad\qquad+2\delta^{\nu}_{\;\;[\rho}  \bar F^{(0)}_{\sigma]\lambda}\bar F^{(1)\mu\lambda}-2\delta^{\mu}_{\;\;[\rho}  \bar F^{(1)}_{\sigma]\lambda}\bar F^{(0)\nu\lambda}-2 g^{\mu\nu} \bar F^{\lambda(0)}_{\;\;\;\;[\rho}\bar F^{(1)}_{\sigma]\lambda}\Big\}a^{(0)}_{\nu}\nonumber\\
&& -2 \sqrt{2} \,h_{\mu\nu}\Big\{\;e^{-2\phi_0} \delta^{(\nu}_{\;\;[\rho} \bar F^{(0)}_{\sigma]\lambda}\bar F^{(0)\mu)\lambda} -\;e^{2\phi_0}\;\delta^{(\nu}_{\;\;[\rho} \bar F^{(1)}_{\sigma]\lambda}\bar F^{(1)\mu)\lambda}\Big\}\phi\nonumber\\
&& -2\sqrt{2}\,\phi \Big\{-\;e^{-2\phi_0} \delta^{(\nu}_{\;\;[\rho} \bar F^{(0)}_{\sigma]\lambda}\bar F^{(0)\mu)\lambda} +\;e^{2\phi_0}\;\delta^{(\nu}_{\;\;[\rho} \bar F^{(1)}_{\sigma]\lambda}\bar F^{(1)\mu)\lambda}\Big\}h_{\mu\nu}
\nonumber\\&&
+h_{\mu\nu}\bigg[\frac{1}{2}\bigg(\delta^{\lambda}_{\;\;[\rho}\delta{}^{(\nu}_{\sigma]}g^{\mu)\kappa}+\delta^{\kappa}_{\;\;[\rho}\delta{}^{(\nu}_{\sigma]}g^{\mu)\lambda}\bigg)\left(e^{-2\phi_0} (F^{(0)})^2+e^{2\phi_0} (F^{(1)})^2-\frac{1}{2} \bar R\right)\nonumber\\&&\qquad\quad+\bigg( \delta^{\lambda}_{\;\;[\rho} \delta^{(\nu}_{\sigma]}\bar R^{\mu)\kappa}+  \delta^{\kappa}_{\;\;[\rho} \delta^{(\nu}_{\sigma]}\bar R^{\mu)\lambda}-g^{\mu\nu}\delta^{(\lambda}_{\;\;[\rho} \bar R^{\;\;\;\kappa)}_{\sigma]} +g^{\lambda\kappa}\delta^{(\nu}_{\;\;[\rho} \bar R^{\;\;\;\mu)}_{\sigma]}\bigg)\nonumber\\&&\qquad\quad+2\bigg\{ e^{-2\phi_0}\bigg(-\bar F^{\kappa(0)}_{\;\;[\rho}\bar F^{(0)(\mu}_{\sigma]}\bar g^{\nu)\lambda}- \bar F^{\lambda(0)}_{\;\;[\rho}\bar F^{(0)(\mu}_{\sigma]}\bar g^{\nu)\kappa}+ \bar F^{(0)\lambda(\mu}\delta^{\nu)}_{\;\;[\rho}\bar F^{(0)\kappa}_{\sigma]} \nonumber\\&&\qquad\quad +\bar F^{(0)\kappa(\mu}\delta^{\nu)}_{\;\;[\rho} \bar F^{(0)\lambda}_{\sigma]}-\bar F^{(0)\nu(\kappa}\delta^{\lambda)}_{\;\;[\rho}\bar F^{(0)\mu}_{\sigma]}-\bar F^{(0)\mu(\kappa}\delta^{\lambda)}_{\;\;[\rho}\bar F^{(0)\nu}_{\sigma]}\bigg)\\&&\qquad\quad+e^{2\phi_0}\bigg(-\bar F^{\kappa(1)}_{\;\;[\rho}\bar F^{(1)(\mu}_{\sigma]}\bar g^{\nu)\lambda}- \bar F^{\lambda(1)}_{\;\;[\rho}\bar F^{(1)(\mu}_{\sigma]}\bar g^{\nu)\kappa}+\bar F^{(1)\lambda(\mu} \delta^{\nu)}_{\;\;[\rho}\bar F^{\kappa(1)}_{\sigma]} \nonumber\\&&\qquad\quad+\bar F^{(1)\kappa(\mu}\delta^{\nu)}_{\;\;[\rho}\bar F^{(1)\lambda}_{\sigma]}-F^{(1)\nu(\kappa}\delta^{\lambda)}_{\;\;[\rho}\bar F^{(1)\mu}_{\sigma]}-F^{(1)\mu(\kappa}\delta^{\lambda)}_{\;\;[\rho}\bar F^{(1)\nu}_{\sigma]}\bigg)\bigg\}\bigg]h_{\lambda\kappa}\nonumber
\end{eqnarray}
Finally the components of ${2}D_{[\rho }\omega_{\sigma]}$ are
\begin{align}
    \begin{split}
       \xi_m ( {2}D_{[\rho }\omega_{\sigma]})^{mn}\xi_n=&\; \;\xi_m (D_{\rho }\omega_{\sigma}-D_{\sigma }\omega_{\rho})^{mn}\xi_n
         \\
       =&\;\;  {\sqrt{2}}\,h_{\mu\nu}\, e^{-\phi_0}\left(D_{\rho}\bar{F}_{\;\sigma}^{(0) (\mu  }\bar{g}^{\nu) \lambda}-D_{\rho} \bar{F}^{(0)\lambda (\mu  }\delta^{\nu)}_{\;\sigma } -\tfrac{1}{2}\bar{g}^{\mu\nu}D_{\rho}\bar{F}_{\;\sigma}^{(0) \lambda}\right)\,a^{(0)}_{\lambda} 
       \\
       & +{\sqrt{2}}\,h_{\mu\nu} \,e^{\phi_0}\left(D_{\rho}\bar{F}_{\;\sigma}^{(1) (\mu  }\bar{g}^{\nu) \lambda }-D_{\rho}\bar{F}^{(1)\lambda (\mu  }\delta^{\nu) }_{\sigma} -\tfrac{1}{2} \bar{g}^{\mu\nu } D_{\rho}\bar{F}_{\;\sigma}^{(1)\lambda} \right)\,a^{(1)}_{\lambda}
       \\
       &- {\sqrt{2}}\,a^{(0)}_{\lambda}  e^{-\phi_0}\left( D_{\rho}\bar{F}_{\sigma}^{(0) (\mu  }\bar{g}^{\nu) \lambda}-D_{\rho}\bar{F}^{(0)\lambda (\mu  }\delta^{\nu)}_{\sigma} -\tfrac{1}{2}\bar{g}^{\mu\nu} D_{\rho}\bar{F}_{\sigma}^{(0) \lambda}\right)h_{\mu\nu}\,
       \\
       & -{\sqrt{2}}\, a^{(1)}_\lambda e^{\phi_0}\left( D_{\rho}\bar{F}_{\sigma}^{(1) (\mu  }\bar{g}^{\nu) \lambda }-D_{\rho}\bar{F}^{(1)\lambda (\mu  }\delta^{\nu)}_{\sigma } - \tfrac{1}{2}\bar{g}^{\mu\nu }D_{\rho}\bar{F}_{\sigma}^{(1)\lambda} \right)h_{\mu\nu}\,
        \\
       & -\,a^{(0)}_\mu  e^{-\phi_0} D_{\rho}\bar{F}_{\;\sigma}^{(0)\mu}\,\phi +\, a^{(1)}_\mu  e^{\phi_0} D_{\rho}\bar{F}_{\;\sigma}^{(1)\mu}\,\phi
       \\
       & + \, \phi\, e^{-\phi_0} D_{\rho}\bar{F}_{\;\sigma}^{(0)\mu} \,a^{(0)}_{\mu} -\, \phi\, e^{\phi_0} D_{\rho}\bar{F}_{\;\sigma}^{(1)\mu}\,a^{(1)}_{\mu}
       \\&-(\rho \leftrightarrow \sigma)\,. 
       \end{split}
\end{align}

Having determined all the necessary building blocks entering the definition of $\Omega_{\rho\sigma}$, we now proceed to evaluate $\mathrm{Tr}\!\left(\Omega_{\rho\sigma}\Omega^{\rho\sigma}\right)$.

\subsection*{For  ${\rm Tr}\,\Omega^2$}
  To obtain the trace for ${\rm \Omega}^2$ we can denote  $\Omega$ as $4\times 4$ matrices in the field space,
  
\begin{equation}
     \qquad (\Omega^2)_m{}^n =\begin{pmatrix}
        \Omega^2_{hh} &\Omega^2_{h0}& \Omega^2_{h1}& \Omega^2_{h\phi}\\
        \Omega^2_{0h}&\Omega^2_{00} &\Omega^2_{01}& \Omega^2_{0\phi}\\
        \Omega^2_{1h}&\Omega^2_{10} &\Omega^2_{11}& \Omega^2_{1\phi}\\
        \Omega^2_{\phi h}&\Omega^2_{\phi 0} &\Omega^2_{\phi1}& \Omega^2_{\phi\phi}   
    \end{pmatrix}\,,
\end{equation}
The trace we want to compute is given by, 
\begin{align}
    \begin{split}
         {\rm Tr} \Omega^2 &= {\rm tr}\Omega^2_{hh} \,+{\rm tr}\Omega^2_{00}\,+{\rm tr}\Omega^2_{11}\, +{\rm tr}\Omega^2_{\phi\phi}
         \\
 &=       {\rm tr}(\Omega_{hh})^2 \,+{\rm tr}(\Omega_{00})^2\,+{\rm tr}(\Omega_{11})^2\, +{\rm tr}(\Omega_{\phi\phi})^2\\
 &\quad +2 {\rm tr}\big( \Omega_{h0}\Omega_{0h }\, +\Omega_{h1}\Omega_{1h }\, + \Omega_{01}\Omega_{10 }\,+\Omega_{0\phi}\Omega_{\phi0 }\,+\,\Omega_{1\phi}\Omega_{\phi1 }\, +\Omega_{h\phi}\Omega_{\phi h }\bigr)\,.
    \end{split}
\end{align}
where the trace of each term is further simplified using the identities \eqref{app:backidentites}, as follows:
\begin{eqnarray}
     {\rm tr} (\Omega_{hh})^2 &=& -12\bar R_{\mu\nu}\bar R^{\mu\nu}+4\bar R\left(e^{-2\phi}(\bar F^{(0)})^2+e^{2\phi}(\bar F^{(1)})^2\right)+3\bar R^2-6 \bar R_{\mu\nu\rho\sigma}\bar R^{\mu\nu\rho\sigma}\,,\nonumber
     \\
     {\rm tr} (\Omega_{00})^2 &=&\!-13e^{-4\phi_0} (\bar{F}^{(0)})^2 (\bar{F}^{(0)})^2  \!+\! e^{-2\phi_0}\!\left( \!8 \bar{R}^{\mu\nu}\bar{F}^{(0)}_{\mu \rho}\bar{F}_{\;\nu}^{(0)\rho} +2 \bar F^{(0)\mu\nu}\bar F^{(0)\rho\sigma}\bar R_{\mu\nu\rho\sigma} \right)\nonumber\\&&\!- \!\bar R_{\mu\nu\rho\sigma} \bar R^{\mu\nu\rho\sigma} \,,\nonumber
      \\
      {\rm tr} (\Omega_{11})^2 &=&-13e^{4\phi_0} (\bar{F}^{(1)})^2 (\bar{F}^{(1)})^2  \!+\! e^{2\phi_0}\!\left( \!8 \bar{R}^{\mu\nu}\bar{F}^{(1)}_{\mu \rho}\bar{F}_{\;\nu}^{(1)\rho} +2 \bar F^{(1)\mu\nu}\bar F^{(1)\rho\sigma}\bar R_{\mu\nu\rho\sigma} \right)\nonumber\\&&\!- \!\bar R_{\mu\nu\rho\sigma} \bar R^{\mu\nu\rho\sigma} \,,\nonumber
      \\
       {\rm tr} (\Omega_{\phi\phi})^2 &=& 0\,,\nonumber
       \\
       2{\rm tr} (\Omega_{h0}E_{0h}) &=& e^{-2\phi_0}\left( -18\bar{R}^{\mu\nu \rho \sigma}\bar{F}^{(0)}_{\mu\nu}\bar{F}^{(0)}_{\rho\sigma} +36  R^{\mu\nu} \bar{F}_{\mu\rho}^{(0)}\bar{F}_{\;\nu}^{(0)\rho}\right)\,,
       \\
       2{\rm tr} (\Omega_{h1}\Omega_{1h}) &=& e^{2\phi_0}\left( -18\bar{R}^{\mu\nu \rho \sigma}\bar{F}^{(1)}_{\mu\nu}\bar{F}^{(1)}_{\rho\sigma} +36 R^{\mu\nu} \bar{F}_{\mu\rho}^{(1)}\bar{F}_{\;\nu}^{(1)\rho}\right)\,,\nonumber
       \\2{\rm tr} (\Omega_{0\phi}\Omega_{\phi 0}) &=& e^{-2\phi_0}\left( -2\bar{R}^{\mu\nu \rho \sigma}\bar{F}^{(0)}_{\mu\nu}\bar{F}^{(0)}_{\rho\sigma} +4 R^{\mu\nu} \bar{F}_{\mu\rho}^{(0)}\bar{F}_{\;\nu}^{(0)\rho}\right)\,,\nonumber
       \\
2{\rm tr} (\Omega_{1\phi}\Omega_{\phi 1}) &=& e^{2\phi_0}\left( -2\bar{R}^{\mu\nu \rho \sigma}\bar{F}^{(1)}_{\mu\nu}\bar{F}^{(1)}_{\rho\sigma} +4 R^{\mu\nu} \bar{F}_{\mu\rho}^{(1)}\bar{F}_{\;\nu}^{(1)\rho}\right)\,,\nonumber      \\ 
        2{\rm tr} (\Omega_{01}\Omega_{10}) &=& -10(\bar{F}^{(0)})^2(\bar{F}^{(1)})^2\,,\nonumber
        \\
         2{\rm tr} (\Omega_{h\phi}\Omega_{\phi h}) &=& 16 (\bar F^{(0)})^2 (\bar F^{(1)})^2-4\bar R_{\mu\nu}\bar R^{\mu\nu}+\bar R^2\,.\nonumber 
\end{eqnarray}
Summing over all the terms, we obtain
\begin{equation}\label{app:traceOmega2}
\begin{aligned}
\text{Tr}(\Omega_{\rho\sigma}\Omega^{\rho\sigma})
&= -8\,\bar R_{\mu\nu\rho\sigma}\bar R^{\mu\nu\rho\sigma}
- 18\Bigl(
e^{-2\phi_0}\,\bar F^{(0)\mu\nu}\bar F^{(0)\lambda\sigma}
+ e^{2\phi_0}\,\bar F^{(1)\mu\nu}\bar F^{(1)\lambda\sigma}
\Bigr)\bar R_{\mu\nu\lambda\sigma}
\\[2pt]
&\quad
- 5\,\bar R_{\mu\nu}\bar R^{\mu\nu}
+ \frac{5}{4}\,\bar R^2
+ 16\,\bar R\Bigl(
e^{-2\phi_0}(\bar F^{(0)})^2
+ e^{2\phi_0}(\bar F^{(1)})^2
\Bigr)
\\[2pt]
&\quad
+ 32\,(\bar F^{(0)})^2(\bar F^{(1)})^2 \, .
\end{aligned}
\end{equation}



\bibliographystyle{JHEP}
\bibliography{ref}

\providecommand{\href}[2]{#2}\begingroup\raggedright\begin{thebibliography}{10}

\bibitem{Sen:2008vm}
A.~Sen, \emph{{Quantum Entropy Function from AdS(2)/CFT(1) Correspondence}},
  \href{https://doi.org/10.1142/S0217751X09045893}{\emph{Int. J. Mod. Phys. A}
  {\bfseries 24} (2009) 4225}
  [\href{https://arxiv.org/abs/0809.3304}{{\ttfamily 0809.3304}}].

\bibitem{Sen_2009}
A.~Sen, \emph{Arithmetic of quantum entropy function},
  \href{https://doi.org/10.1088/1126-6708/2009/08/068}{\emph{Journal of High
  Energy Physics} {\bfseries 2009} (2009) 068–068}.

\bibitem{Gibbons:1976ue}
G.~W. Gibbons and S.~W. Hawking, \emph{{Action Integrals and Partition
  Functions in Quantum Gravity}},
  \href{https://doi.org/10.1103/PhysRevD.15.2752}{\emph{Phys. Rev. D}
  {\bfseries 15} (1977) 2752}.

\bibitem{Hawking:1978jz}
S.~W. Hawking, \emph{{Quantum Gravity and Path Integrals}},
  \href{https://doi.org/10.1103/PhysRevD.18.1747}{\emph{Phys. Rev. D}
  {\bfseries 18} (1978) 1747}.

\bibitem{Sen:2008yk}
A.~Sen, \emph{{Entropy Function and AdS(2) / CFT(1) Correspondence}},
  \href{https://doi.org/10.1088/1126-6708/2008/11/075}{\emph{JHEP} {\bfseries
  11} (2008) 075} [\href{https://arxiv.org/abs/0805.0095}{{\ttfamily
  0805.0095}}].

\bibitem{Bekenstein:1973ur}
J.~D. Bekenstein, \emph{{Black holes and entropy}},
  \href{https://doi.org/10.1103/PhysRevD.7.2333}{\emph{Phys. Rev. D} {\bfseries
  7} (1973) 2333}.

\bibitem{Hawking:1975vcx}
S.~W. Hawking, \emph{{Particle Creation by Black Holes}},
  \href{https://doi.org/10.1007/BF02345020}{\emph{Commun. Math. Phys.}
  {\bfseries 43} (1975) 199}.

\bibitem{Banerjee:2010qc}
S.~Banerjee, R.~K. Gupta and A.~Sen, \emph{{Logarithmic Corrections to Extremal
  Black Hole Entropy from Quantum Entropy Function}},
  \href{https://doi.org/10.1007/JHEP03(2011)147}{\emph{JHEP} {\bfseries 03}
  (2011) 147} [\href{https://arxiv.org/abs/1005.3044}{{\ttfamily 1005.3044}}].

\bibitem{Banerjee:2011jp}
S.~Banerjee, R.~K. Gupta, I.~Mandal and A.~Sen, \emph{{Logarithmic Corrections
  to N=4 and N=8 Black Hole Entropy: A One Loop Test of Quantum Gravity}},
  \href{https://doi.org/10.1007/JHEP11(2011)143}{\emph{JHEP} {\bfseries 11}
  (2011) 143} [\href{https://arxiv.org/abs/1106.0080}{{\ttfamily 1106.0080}}].

\bibitem{Sen:2012kpz}
A.~Sen, \emph{{Logarithmic Corrections to N=2 Black Hole Entropy: An Infrared
  Window into the Microstates}},
  \href{https://doi.org/10.1007/s10714-012-1336-5}{\emph{Gen. Rel. Grav.}
  {\bfseries 44} (2012) 1207}
  [\href{https://arxiv.org/abs/1108.3842}{{\ttfamily 1108.3842}}].

\bibitem{Sen:2012cj}
A.~Sen, \emph{{Logarithmic Corrections to Rotating Extremal Black Hole Entropy
  in Four and Five Dimensions}},
  \href{https://doi.org/10.1007/s10714-012-1373-0}{\emph{Gen. Rel. Grav.}
  {\bfseries 44} (2012) 1947}
  [\href{https://arxiv.org/abs/1109.3706}{{\ttfamily 1109.3706}}].

\bibitem{Sen:2014aja}
A.~Sen, \emph{{Microscopic and Macroscopic Entropy of Extremal Black Holes in
  String Theory}}, \href{https://doi.org/10.1007/s10714-014-1711-5}{\emph{Gen.
  Rel. Grav.} {\bfseries 46} (2014) 1711}
  [\href{https://arxiv.org/abs/1402.0109}{{\ttfamily 1402.0109}}].

\bibitem{bhattacharyya2013onelooptestquantumsupergravity}
S.~Bhattacharyya, A.~Grassi, M.~Marino and A.~Sen, \emph{A one-loop test of
  quantum supergravity},  2013.

\bibitem{Bobev:2023dwx}
N.~Bobev, M.~David, J.~Hong, V.~Reys and X.~Zhang, \emph{{A compendium of
  logarithmic corrections in AdS/CFT}},
  \href{https://doi.org/10.1007/JHEP04(2024)020}{\emph{JHEP} {\bfseries 04}
  (2024) 020} [\href{https://arxiv.org/abs/2312.08909}{{\ttfamily
  2312.08909}}].

\bibitem{Jeon_2017}
I.~Jeon and S.~Lal, \emph{Logarithmic corrections to entropy of magnetically
  charged ads 4 black holes},
  \href{https://doi.org/10.1016/j.physletb.2017.09.026}{\emph{Physics Letters
  B} {\bfseries 774} (2017) 41–45}.

\bibitem{Liu:2017vll}
J.~T. Liu, L.~A. Pando~Zayas, V.~Rathee and W.~Zhao, \emph{{Toward Microstate
  Counting Beyond Large N in Localization and the Dual One-loop Quantum
  Supergravity}}, \href{https://doi.org/10.1007/JHEP01(2018)026}{\emph{JHEP}
  {\bfseries 01} (2018) 026}
  [\href{https://arxiv.org/abs/1707.04197}{{\ttfamily 1707.04197}}].

\bibitem{Liu:2017vbl}
J.~T. Liu, L.~A. Pando~Zayas, V.~Rathee and W.~Zhao, \emph{{One-Loop Test of
  Quantum Black Holes in anti{\textendash}de Sitter Space}},
  \href{https://doi.org/10.1103/PhysRevLett.120.221602}{\emph{Phys. Rev. Lett.}
  {\bfseries 120} (2018) 221602}
  [\href{https://arxiv.org/abs/1711.01076}{{\ttfamily 1711.01076}}].

\bibitem{PandoZayas:2020iqr}
L.~A. Pando~Zayas and Y.~Xin, \emph{{Universal logarithmic behavior in
  microstate counting and the dual one-loop entropy of $AdS_4$ black holes}},
  \href{https://doi.org/10.1103/PhysRevD.103.026003}{\emph{Phys. Rev. D}
  {\bfseries 103} (2021) 026003}
  [\href{https://arxiv.org/abs/2008.03239}{{\ttfamily 2008.03239}}].

\bibitem{Gang:2019uay}
D.~Gang, N.~Kim and L.~A. Pando~Zayas, \emph{{Precision Microstate Counting for
  the Entropy of Wrapped M5-branes}},
  \href{https://doi.org/10.1007/JHEP03(2020)164}{\emph{JHEP} {\bfseries 03}
  (2020) 164} [\href{https://arxiv.org/abs/1905.01559}{{\ttfamily
  1905.01559}}].

\bibitem{Benini:2019dyp}
F.~Benini, D.~Gang and L.~A. Pando~Zayas, \emph{{Rotating Black Hole Entropy
  from M5 Branes}}, \href{https://doi.org/10.1007/JHEP03(2020)057}{\emph{JHEP}
  {\bfseries 03} (2020) 057}
  [\href{https://arxiv.org/abs/1909.11612}{{\ttfamily 1909.11612}}].

\bibitem{David:2021eoq}
M.~David, V.~Godet, Z.~Liu and L.~A. Pando~Zayas, \emph{{Non-topological
  logarithmic corrections in minimal gauged supergravity}},
  \href{https://doi.org/10.1007/JHEP08(2022)043}{\emph{JHEP} {\bfseries 08}
  (2022) 043} [\href{https://arxiv.org/abs/2112.09444}{{\ttfamily
  2112.09444}}].

\bibitem{Karan:2022dfy}
S.~Karan and G.~S. Punia, \emph{{Logarithmic correction to black hole entropy
  in universal low-energy string theory models}},
  \href{https://doi.org/10.1007/JHEP03(2023)028}{\emph{JHEP} {\bfseries 03}
  (2023) 028} [\href{https://arxiv.org/abs/2210.16230}{{\ttfamily
  2210.16230}}].

\bibitem{Karan:2024gwf}
S.~Karan, G.~S. Punia and S.~Biswas, \emph{{Logarithmic correction to the
  entropy of a Kerr-Newman family of black holes in U(1)2-charged STU
  supergravity models}},
  \href{https://doi.org/10.1103/PhysRevD.111.066016}{\emph{Phys. Rev. D}
  {\bfseries 111} (2025) 066016}
  [\href{https://arxiv.org/abs/2403.11823}{{\ttfamily 2403.11823}}].

\bibitem{Sen:2012dw}
A.~Sen, \emph{{Logarithmic Corrections to Schwarzschild and Other Non-extremal
  Black Hole Entropy in Different Dimensions}},
  \href{https://doi.org/10.1007/JHEP04(2013)156}{\emph{JHEP} {\bfseries 04}
  (2013) 156} [\href{https://arxiv.org/abs/1205.0971}{{\ttfamily 1205.0971}}].

\bibitem{Banerjee:2021pdy}
G.~Banerjee and B.~Panda, \emph{{Logarithmic corrections to the entropy of
  non-extremal black holes in $ \mathcal{N} $ = 1 Einstein-Maxwell
  supergravity}}, \href{https://doi.org/10.1007/JHEP11(2021)214}{\emph{JHEP}
  {\bfseries 11} (2021) 214}
  [\href{https://arxiv.org/abs/2109.04407}{{\ttfamily 2109.04407}}].

\bibitem{Charles:2015eha}
A.~M. Charles and F.~Larsen, \emph{{Universal corrections to non-extremal black
  hole entropy in $ \mathcal{N}\ge 2 $ supergravity}},
  \href{https://doi.org/10.1007/JHEP06(2015)200}{\emph{JHEP} {\bfseries 06}
  (2015) 200} [\href{https://arxiv.org/abs/1505.01156}{{\ttfamily
  1505.01156}}].

\bibitem{Bhattacharyya:2012wz}
S.~Bhattacharyya, B.~Panda and A.~Sen, \emph{{Heat Kernel Expansion and
  Extremal Kerr-Newmann Black Hole Entropy in Einstein-Maxwell Theory}},
  \href{https://doi.org/10.1007/JHEP08(2012)084}{\emph{JHEP} {\bfseries 08}
  (2012) 084} [\href{https://arxiv.org/abs/1204.4061}{{\ttfamily 1204.4061}}].

\bibitem{Castro:2018hsc}
A.~Castro, V.~Godet, F.~Larsen and Y.~Zeng, \emph{{Logarithmic Corrections to
  Black Hole Entropy: the Non-BPS Branch}},
  \href{https://doi.org/10.1007/JHEP05(2018)079}{\emph{JHEP} {\bfseries 05}
  (2018) 079} [\href{https://arxiv.org/abs/1801.01926}{{\ttfamily
  1801.01926}}].

\bibitem{Karan:2019gyn}
S.~Karan, G.~Banerjee and B.~Panda, \emph{{Seeley-DeWitt Coefficients in
  $\mathcal{N}=2$ Einstein-Maxwell Supergravity Theory and Logarithmic
  Corrections to $\mathcal{N}=2$ Extremal Black Hole Entropy}},
  \href{https://doi.org/10.1007/JHEP08(2019)056}{\emph{JHEP} {\bfseries 08}
  (2019) 056} [\href{https://arxiv.org/abs/1905.13058}{{\ttfamily
  1905.13058}}].

\bibitem{Karan:2020njm}
S.~Karan and B.~Panda, \emph{{Logarithmic corrections to black hole entropy in
  matter coupled $\mathcal{N} \geq 1$ Einstein-Maxwell supergravity}},
  \href{https://doi.org/10.1007/JHEP05(2021)104}{\emph{JHEP} {\bfseries 05}
  (2021) 104} [\href{https://arxiv.org/abs/2012.12227}{{\ttfamily
  2012.12227}}].

\bibitem{Karan:2021teq}
S.~Karan and B.~Panda, \emph{{Generalized Einstein-Maxwell theory:
  Seeley-DeWitt coefficients and logarithmic corrections to the entropy of
  extremal and nonextremal black holes}},
  \href{https://doi.org/10.1103/PhysRevD.104.046010}{\emph{Phys. Rev. D}
  {\bfseries 104} (2021) 046010}
  [\href{https://arxiv.org/abs/2104.06381}{{\ttfamily 2104.06381}}].

\bibitem{Iliesiu:2021are}
L.~V. Iliesiu, M.~Kologlu and G.~J. Turiaci, \emph{{Supersymmetric indices
  factorize}}, \href{https://doi.org/10.1007/JHEP05(2023)032}{\emph{JHEP}
  {\bfseries 05} (2023) 032}
  [\href{https://arxiv.org/abs/2107.09062}{{\ttfamily 2107.09062}}].

\bibitem{H:2023qko}
A.~A. H., P.~V. Athira, C.~Chowdhury and A.~Sen, \emph{{Logarithmic correction
  to BPS black hole entropy from supersymmetric index at finite temperature}},
  \href{https://doi.org/10.1007/JHEP03(2024)095}{\emph{JHEP} {\bfseries 03}
  (2024) 095} [\href{https://arxiv.org/abs/2306.07322}{{\ttfamily
  2306.07322}}].

\bibitem{Anupam:2023yns}
A.~H. Anupam, C.~Chowdhury and A.~Sen, \emph{{Revisiting logarithmic correction
  to five dimensional BPS black hole entropy}},
  \href{https://doi.org/10.1007/JHEP05(2024)070}{\emph{JHEP} {\bfseries 05}
  (2024) 070} [\href{https://arxiv.org/abs/2308.00038}{{\ttfamily
  2308.00038}}].

\bibitem{Iliesiu:2022onk}
L.~V. Iliesiu, S.~Murthy and G.~J. Turiaci, \emph{{Revisiting the logarithmic
  corrections to the black hole entropy}},
  \href{https://doi.org/10.1007/JHEP07(2025)058}{\emph{JHEP} {\bfseries 07}
  (2025) 058} [\href{https://arxiv.org/abs/2209.13608}{{\ttfamily
  2209.13608}}].

\bibitem{Banerjee:2023quv}
N.~Banerjee and M.~Saha, \emph{{Revisiting leading quantum corrections to near
  extremal black hole thermodynamics}},
  \href{https://doi.org/10.1007/JHEP07(2023)010}{\emph{JHEP} {\bfseries 07}
  (2023) 010} [\href{https://arxiv.org/abs/2303.12415}{{\ttfamily
  2303.12415}}].

\bibitem{Banerjee:2023gll}
N.~Banerjee, M.~Saha and S.~Srinivasan, \emph{{Logarithmic corrections for
  near-extremal black holes}},
  \href{https://doi.org/10.1007/JHEP02(2024)077}{\emph{JHEP} {\bfseries 2024}
  (2024) 077} [\href{https://arxiv.org/abs/2311.09595}{{\ttfamily
  2311.09595}}].

\bibitem{Modak:2025gvp}
A.~Modak, A.~Singh and B.~Panda, \emph{{Logarithmic Corrections for
  Near-extremal Kerr-Newman Black Holes}},
  \href{https://arxiv.org/abs/2502.18173}{{\ttfamily 2502.18173}}.

\bibitem{Caldarelli:1998hg}
M.~M. Caldarelli and D.~Klemm, \emph{{Supersymmetry of Anti-de Sitter black
  holes}}, \href{https://doi.org/10.1016/S0550-3213(98)00846-3}{\emph{Nucl.
  Phys. B} {\bfseries 545} (1999) 434}
  [\href{https://arxiv.org/abs/hep-th/9808097}{{\ttfamily hep-th/9808097}}].

\bibitem{Cacciatori:2008ek}
S.~L. Cacciatori, D.~Klemm, D.~S. Mansi and E.~Zorzan, \emph{{All timelike
  supersymmetric solutions of N=2, D=4 gauged supergravity coupled to abelian
  vector multiplets}},
  \href{https://doi.org/10.1088/1126-6708/2008/05/097}{\emph{JHEP} {\bfseries
  05} (2008) 097} [\href{https://arxiv.org/abs/0804.0009}{{\ttfamily
  0804.0009}}].

\bibitem{Gutowski:2004ez}
J.~B. Gutowski and H.~S. Reall, \emph{{Supersymmetric AdS(5) black holes}},
  \href{https://doi.org/10.1088/1126-6708/2004/02/006}{\emph{JHEP} {\bfseries
  02} (2004) 006} [\href{https://arxiv.org/abs/hep-th/0401042}{{\ttfamily
  hep-th/0401042}}].

\bibitem{Cacciatori:2009iz}
S.~L. Cacciatori and D.~Klemm, \emph{{Supersymmetric AdS(4) black holes and
  attractors}}, \href{https://doi.org/10.1007/JHEP01(2010)085}{\emph{JHEP}
  {\bfseries 01} (2010) 085} [\href{https://arxiv.org/abs/0911.4926}{{\ttfamily
  0911.4926}}].

\bibitem{Hristov:2010ri}
K.~Hristov and S.~Vandoren, \emph{{Static supersymmetric black holes in
  AdS$_{4}$ with spherical symmetry}},
  \href{https://doi.org/10.1007/JHEP04(2011)047}{\emph{JHEP} {\bfseries 04}
  (2011) 047} [\href{https://arxiv.org/abs/1012.4314}{{\ttfamily 1012.4314}}].

\bibitem{Benini:2015eyy}
F.~Benini, K.~Hristov and A.~Zaffaroni, \emph{{Black hole microstates in
  AdS$_{4}$ from supersymmetric localization}},
  \href{https://doi.org/10.1007/JHEP05(2016)054}{\emph{JHEP} {\bfseries 05}
  (2016) 054} [\href{https://arxiv.org/abs/1511.04085}{{\ttfamily
  1511.04085}}].

\bibitem{Cabo-Bizet:2018ehj}
A.~Cabo-Bizet, D.~Cassani, D.~Martelli and S.~Murthy, \emph{{Microscopic origin
  of the Bekenstein-Hawking entropy of supersymmetric AdS$_{5}$ black holes}},
  \href{https://doi.org/10.1007/JHEP10(2019)062}{\emph{JHEP} {\bfseries 10}
  (2019) 062} [\href{https://arxiv.org/abs/1810.11442}{{\ttfamily
  1810.11442}}].

\bibitem{Choi:2018hmj}
S.~Choi, J.~Kim, S.~Kim and J.~Nahmgoong, \emph{{Large AdS black holes from
  QFT}},  \href{https://arxiv.org/abs/1810.12067}{{\ttfamily 1810.12067}}.

\bibitem{Zaffaroni:2019dhb}
A.~Zaffaroni, \emph{{AdS black holes, holography and localization}},
  \href{https://doi.org/10.1007/s41114-020-00027-8}{\emph{Living Rev. Rel.}
  {\bfseries 23} (2020) 2} [\href{https://arxiv.org/abs/1902.07176}{{\ttfamily
  1902.07176}}].

\bibitem{Cabo-Bizet:2017jsl}
A.~Cabo-Bizet, V.~I. Giraldo-Rivera and L.~A. Pando~Zayas, \emph{{Microstate
  counting of AdS$_{4}$ hyperbolic black hole entropy via the topologically
  twisted index}}, \href{https://doi.org/10.1007/JHEP08(2017)023}{\emph{JHEP}
  {\bfseries 08} (2017) 023}
  [\href{https://arxiv.org/abs/1701.07893}{{\ttfamily 1701.07893}}].

\bibitem{Ferrara_1995}
S.~Ferrara, R.~Kallosh and A.~Strominger, \emph{N=2 extremal black holes},
  \href{https://doi.org/10.1103/physrevd.52.r5412}{\emph{Physical Review D}
  {\bfseries 52} (1995) R5412–R5416}.

\bibitem{Ferrara:1996dd}
S.~Ferrara and R.~Kallosh, \emph{{Supersymmetry and attractors}},
  \href{https://doi.org/10.1103/PhysRevD.54.1514}{\emph{Phys. Rev. D}
  {\bfseries 54} (1996) 1514}
  [\href{https://arxiv.org/abs/hep-th/9602136}{{\ttfamily hep-th/9602136}}].

\bibitem{Ferrara:1996um}
S.~Ferrara and R.~Kallosh, \emph{{Universality of supersymmetric attractors}},
  \href{https://doi.org/10.1103/PhysRevD.54.1525}{\emph{Phys. Rev. D}
  {\bfseries 54} (1996) 1525}
  [\href{https://arxiv.org/abs/hep-th/9603090}{{\ttfamily hep-th/9603090}}].

\bibitem{Strominger_1996}
A.~Strominger, \emph{Macroscopic entropy of n = 2 extremal black holes},
  \href{https://doi.org/10.1016/0370-2693(96)00711-3}{\emph{Physics Letters B}
  {\bfseries 383} (1996) 39–43}.

\bibitem{Ferrara_1997}
S.~Ferrara, G.~W. Gibbons and R.~Kallosh, \emph{Black holes and critical points
  in moduli space},
  \href{https://doi.org/10.1016/s0550-3213(97)00324-6}{\emph{Nuclear Physics B}
  {\bfseries 500} (1997) 75–93}.

\bibitem{Vassilevich:2003xt}
D.~V. Vassilevich, \emph{{Heat kernel expansion: User's manual}},
  \href{https://doi.org/10.1016/j.physrep.2003.09.002}{\emph{Phys. Rept.}
  {\bfseries 388} (2003) 279}
  [\href{https://arxiv.org/abs/hep-th/0306138}{{\ttfamily hep-th/0306138}}].

\bibitem{Avramidi:1994fx}
I.~G. Avramidi, \emph{{The Heat kernel approach for calculating the effective
  action in quantum field theory and quantum gravity}},
  \href{https://arxiv.org/abs/hep-th/9509077}{{\ttfamily hep-th/9509077}}.

\bibitem{Hawking:1976ja}
S.~W. Hawking, \emph{{Zeta Function Regularization of Path Integrals in Curved
  Space-Time}}, \href{https://doi.org/10.1007/BF01626516}{\emph{Commun. Math.
  Phys.} {\bfseries 55} (1977) 133}.

\bibitem{Denardo:1982mf}
G.~Denardo and E.~Spallucci, \emph{{Induced Quantum Gravity From Heat Kernel
  Expansion}}, \href{https://doi.org/10.1007/BF02902652}{\emph{Nuovo Cim. A}
  {\bfseries 69} (1982) 151}.

\bibitem{Barvinsky:2015bky}
A.~Barvinsky, \emph{{Heat kernel expansion in the background field formalism}},
  \href{https://doi.org/10.4249/scholarpedia.31644}{\emph{Scholarpedia}
  {\bfseries 10} (2015) 31644}.

\bibitem{CHRISTENSEN1980480}
S.~Christensen and M.~Duff, \emph{Quantizing gravity with a cosmological
  constant},
  \href{https://doi.org/https://doi.org/10.1016/0550-3213(80)90423-X}{\emph{Nuclear
  Physics B} {\bfseries 170} (1980) 480}.

\bibitem{Hristov:2018lod}
K.~Hristov, I.~Lodato and V.~Reys, \emph{{On the quantum entropy function in 4d
  gauged supergravity}},
  \href{https://doi.org/10.1007/JHEP07(2018)072}{\emph{JHEP} {\bfseries 07}
  (2018) 072} [\href{https://arxiv.org/abs/1803.05920}{{\ttfamily
  1803.05920}}].

\bibitem{Hristov:2019xku}
K.~Hristov, I.~Lodato and V.~Reys, \emph{{One-loop determinants for black holes
  in 4d gauged supergravity}},
  \href{https://doi.org/10.1007/JHEP11(2019)105}{\emph{JHEP} {\bfseries 11}
  (2019) 105} [\href{https://arxiv.org/abs/1908.05696}{{\ttfamily
  1908.05696}}].

\bibitem{Dabholkar:2010uh}
A.~Dabholkar, J.~Gomes and S.~Murthy, \emph{{Quantum black holes, localization
  and the topological string}},
  \href{https://doi.org/10.1007/JHEP06(2011)019}{\emph{JHEP} {\bfseries 06}
  (2011) 019} [\href{https://arxiv.org/abs/1012.0265}{{\ttfamily 1012.0265}}].

\bibitem{Dabholkar:2011ec}
A.~Dabholkar, J.~Gomes and S.~Murthy, \emph{{Localization \& Exact
  Holography}}, \href{https://doi.org/10.1007/JHEP04(2013)062}{\emph{JHEP}
  {\bfseries 04} (2013) 062} [\href{https://arxiv.org/abs/1111.1161}{{\ttfamily
  1111.1161}}].

\bibitem{Jeon:2018kec}
I.~Jeon and S.~Murthy, \emph{{Twisting and localization in supergravity:
  equivariant cohomology of BPS black holes}},
  \href{https://doi.org/10.1007/JHEP03(2019)140}{\emph{JHEP} {\bfseries 03}
  (2019) 140} [\href{https://arxiv.org/abs/1806.04479}{{\ttfamily
  1806.04479}}].

\bibitem{Iliesiu:2022kny}
L.~V. Iliesiu, S.~Murthy and G.~J. Turiaci, \emph{{Black hole microstate
  counting from the gravitational path integral}},
  \href{https://arxiv.org/abs/2209.13602}{{\ttfamily 2209.13602}}.

\bibitem{Dabholkar:2014wpa}
A.~Dabholkar, N.~Drukker and J.~Gomes, \emph{{Localization in supergravity and
  quantum $AdS_4/CFT_3$ holography}},
  \href{https://doi.org/10.1007/JHEP10(2014)090}{\emph{JHEP} {\bfseries 10}
  (2014) 090} [\href{https://arxiv.org/abs/1406.0505}{{\ttfamily 1406.0505}}].

\bibitem{Nian:2017hac}
J.~Nian and X.~Zhang, \emph{{Entanglement Entropy of ABJM Theory and Entropy of
  Topological Black Hole}},
  \href{https://doi.org/10.1007/JHEP07(2017)096}{\emph{JHEP} {\bfseries 07}
  (2017) 096} [\href{https://arxiv.org/abs/1705.01896}{{\ttfamily
  1705.01896}}].

\bibitem{Hristov:2021zai}
K.~Hristov and V.~Reys, \emph{{Factorization of log-corrections in
  AdS$_{4}$/CFT$_{3}$ from supergravity localization}},
  \href{https://doi.org/10.1007/JHEP12(2021)031}{\emph{JHEP} {\bfseries 12}
  (2021) 031} [\href{https://arxiv.org/abs/2107.12398}{{\ttfamily
  2107.12398}}].

\bibitem{LopesCardoso:2022hvc}
G.~Lopes~Cardoso, A.~Kidambi, S.~Nampuri, V.~Reys and M.~Rossell\'o, \emph{{The
  gravitational path integral for $ N=4$ BPS black holes from black hole
  microstate counting}},  \href{https://arxiv.org/abs/2211.06873}{{\ttfamily
  2211.06873}}.

\bibitem{GonzalezLezcano:2023cuh}
A.~Gonz\'alez~Lezcano, I.~Jeon and A.~Ray, \emph{{Supersymmetric localization:
  ${\cal N}=(2,2)$ theories on S$^2$ and AdS$_2$}},
  \href{https://arxiv.org/abs/2302.10370}{{\ttfamily 2302.10370}}.

\bibitem{GonzalezLezcano:2023uar}
A.~Gonz{\'a}lez~Lezcano, I.~Jeon and A.~Ray, \emph{{Supersymmetry and
  complexified spectrum on Euclidean AdS2}},
  \href{https://doi.org/10.1103/PhysRevD.108.045018}{\emph{Phys. Rev. D}
  {\bfseries 108} (2023) 045018}
  [\href{https://arxiv.org/abs/2305.12925}{{\ttfamily 2305.12925}}].

\bibitem{GonzalezLezcano:2024rsi}
A.~Gonz{\'a}lez~Lezcano, I.~Jeon and A.~Ray, \emph{{Supersymmetric spectrum for
  vector multiplet on Euclidean AdS$_{2}$}},
  \href{https://doi.org/10.1007/JHEP08(2024)139}{\emph{JHEP} {\bfseries 08}
  (2024) 139} [\href{https://arxiv.org/abs/2404.18376}{{\ttfamily
  2404.18376}}].

\bibitem{Cacciatori:2004rt}
S.~L. Cacciatori, M.~M. Caldarelli, D.~Klemm and D.~S. Mansi, \emph{{More on
  BPS solutions of N = 2, D = 4 gauged supergravity}},
  \href{https://doi.org/10.1088/1126-6708/2004/07/061}{\emph{JHEP} {\bfseries
  07} (2004) 061} [\href{https://arxiv.org/abs/hep-th/0406238}{{\ttfamily
  hep-th/0406238}}].

\bibitem{VanProeyen:2004xt}
A.~Van~Proeyen, \emph{{Supergravity with Fayet-Iliopoulos terms and
  R-symmetry}}, \href{https://doi.org/10.1002/prop.200410248}{\emph{Fortsch.
  Phys.} {\bfseries 53} (2005) 997}
  [\href{https://arxiv.org/abs/hep-th/0410053}{{\ttfamily hep-th/0410053}}].

\bibitem{Freedman:2012zz}
D.~Z. Freedman and A.~Van~Proeyen, \emph{{Supergravity}}. Cambridge Univ.
  Press, Cambridge, UK, 5, 2012,
  \href{https://doi.org/10.1017/CBO9781139026833}{10.1017/CBO9781139026833}.

\bibitem{Astesiano:2020fwe}
D.~Astesiano and S.~L. Cacciatori, \emph{{Super throats with non trivial
  scalars}}, \href{https://doi.org/10.1007/JHEP07(2020)017}{\emph{JHEP}
  {\bfseries 07} (2020) 017}
  [\href{https://arxiv.org/abs/2003.11582}{{\ttfamily 2003.11582}}].

\bibitem{Chimento:2015rra}
S.~Chimento, D.~Klemm and N.~Petri, \emph{{Supersymmetric black holes and
  attractors in gauged supergravity with hypermultiplets}},
  \href{https://doi.org/10.1007/JHEP06(2015)150}{\emph{JHEP} {\bfseries 06}
  (2015) 150} [\href{https://arxiv.org/abs/1503.09055}{{\ttfamily
  1503.09055}}].

\bibitem{Bellucci:2008cb}
S.~Bellucci, S.~Ferrara, A.~Marrani and A.~Yeranyan, \emph{{d=4 Black Hole
  Attractors in N=2 Supergravity with Fayet-Iliopoulos Terms}},
  \href{https://doi.org/10.1103/PhysRevD.77.085027}{\emph{Phys. Rev. D}
  {\bfseries 77} (2008) 085027}
  [\href{https://arxiv.org/abs/0802.0141}{{\ttfamily 0802.0141}}].

\end{thebibliography}\endgroup
\end{document}